\documentclass{emulateapj}

\usepackage{natbib} 
\begin{document}

\def\mpch {$h^{-1}$ Mpc} 
\def\kms {km s$^{-1}$} 
\def\lcdm {$\Lambda$CDM } 
\def\wt {$\omega(\theta)$}
\def\Aw {$A_{\omega}$}
\def\onew {$A_{\omega}(1\arcmin)$}
\def\xir {$\xi(r)$}
\def\xix {$\xi(x)$}

\title{Evolution and Color-Dependence of the Galaxy Angular Correlation 
Function: 350,000 Galaxies in 5 Square Degrees}
\author{Alison L. Coil\altaffilmark{1}, 
Jeffrey A. Newman\altaffilmark{2}, Nick Kaiser\altaffilmark{3}, 
Marc Davis\altaffilmark{1}, Chung-Pei Ma\altaffilmark{1}, 
Dale D. Kocevski\altaffilmark{3}, David C. Koo\altaffilmark{4}}
\altaffiltext{1}{Department of Astronomy, University of California,
Berkeley, CA 94720 -- 3411} \altaffiltext{2}{Hubble Fellow; Institute for 
Nuclear and Particle Astrophysics, Lawrence Berkeley National Laboratory, 
Berkeley, CA 94720} \altaffiltext{3}{Institute for Astronomy,
University of Hawaii, 2680 Woodlawn Drive, Honolulu, HI 96822}
\altaffiltext{4}{University of California Observatories/Lick
Observatory, Department of Astronomy and Astrophysics, University of
California, Santa Cruz, CA 95064}

\slugcomment{Accepted for publication by ApJ}

\begin{abstract}

When applied to deep 
photometric catalogs, the two-point angular correlation function, \wt, 
is a sensitive probe of the evolution of 
galaxy clustering properties.  Here we present measurements of 
\wt \ as a function of $I_{\rm AB}$ magnitude 
and $(R-I)$ color to a depth of $I_{\rm AB}=24$ on scales
$\theta=7\arcsec -3\arcmin$, 
using a sample of $\sim350,000$ galaxies covering 5 degrees$^2$ in
total over 5 separate fields.  Using redshifts 
of 2954 galaxies in early DEEP2 Galaxy Redshift Survey data, we
construct robust galaxy redshift distributions as a
function of $I_{\rm AB}$ and $R_{\rm AB}$ magnitude and $(R-I)$ 
color for galaxies between
$0<z<2$.  We constrain models of the redshift evolution of galaxy 
clustering and find that significant growth of clustering has occurred 
from $z\geq1$ to $z=0$.  A model in which the comoving scale-length, $x_0$,
evolves linearly with redshift, $x_0(z)=x_0(0)(1-Bz)$, fits the data
better than the $\epsilon$ model proposed by \cite{Groth77}.
The clustering properties depend strongly on observed $(R-I)$
color, with both the reddest and bluest galaxies exhibiting large
clustering amplitudes and steeper slopes. 
Different observed $(R-I)$ color ranges are sensitive to very 
disparate redshift regimes.  Red galaxies with $(R-I)\sim1.5$ lie in a narrow
redshift range centered at $z\sim0.85$ and have a comoving scale
length of clustering at $z=0.85$ of $x_0=5.0 \pm0.3$ \mpch.  These 
galaxies have early-type spectra and are likely progenitors of
massive local ellipticals.  The bluest galaxies with $(R-I)\sim0$ appear
to be a mix of star-forming galaxies, both relatively local ($z\sim0.3-0.6$) 
dwarfs and bright $z>1.4$ galaxies, and broad-line AGN.   
We find that local blue dwarfs are relatively unclustered, 
with $x_0=1.6 \pm0.2$ \mpch.
The $z>1.4$ blue galaxies have a larger clustering scale-length,
$x_0\gtrsim$5 \mpch. 

\end{abstract}

\keywords{galaxies: statistics --- galaxies: evolution ---
cosmology: large-scale structure of universe}

\section{Introduction}

Studies of galaxy clustering can constrain cosmological parameters as
well as galaxy formation models, since the observed galaxy clustering
strength depends on both the underlying dark matter distribution and
the efficiency of galaxy formation inside dark matter halos.  While
theory and simulations both provide detailed predictions for the dark
matter distribution in various cosmological models, the processes of
galaxy formation and evolution are less well-understood.  The redshift
dependence of the galaxy clustering strength, in particular, can be
used to constrain galaxy evolution theories if the cosmology is known.

Galaxy clustering is ideally studied in three dimensions, where the
redshift of each galaxy is known.  However, cosmic variance dictates
the need for very large data samples, probing a large volume of space
and sampling multiple independent fields, in order to obtain
statistically robust results.  As photometric data exist over larger
areas and to fainter magnitude limits than are currently available for
spectroscopic samples, galaxy clustering has long been measured from
two-dimensional photometry alone.

The angular two-point correlation function, \wt, measures the excess
probability above random of finding a galaxy at a specific angle
$\theta$ from another galaxy.  This quantity is derived from counts of
galaxy pairs as a function of their separation, and while relatively
straightforward to measure, its interpretation is not trivial.  From
the projected angular two-point correlation function, \wt, one can
infer the three-dimensional spatial two-point correlation function,
$\xi(r)$, if the redshift distribution of the sources in each sample
is well-known and the redshift dependence of the clustering strength
is known or assumed.  The two-point correlation function, $\xi(r)$, 
is usually fit as a power law,
$\xi(r)=(r/r_0)^{-\gamma}$, where $r_0$ is the scale-length of the
galaxy clustering.  However, the assumed galaxy redshift distribution
($dN/dz$) has varied widely in different studies, such that similar
observed angular clustering results have led to widely different
conclusions.  A further complication is that each sample usually spans
a large range of redshifts and is magnitude-limited, such that the
mean intrinsic luminosity of the galaxies is changing with redshift
within a sample.  As the clustering properties of galaxies are expected
to depend on luminosity and have been shown to do so locally  
\citep{Norberg01}, this can hinder
interpretation of the evolution of clustering measured in \wt \ studies.

Many of the first useful measurements of large-scale structure were
studies of angular clustering. 
One of the earliest measurements of the \wt \ statistic was the
pioneering work of \cite{Peebles75} using photographic plates from the
Lick survey.  They found \wt \ to be well fit by a power law,
\wt$=A_{\omega}\theta^{\delta}$, with a slope of $\delta=-0.77 \pm0.06$.  
Later
studies using CCDs were able to reach deeper magnitude limits and
found that fainter galaxies had a lower clustering amplitude.
However, most of these surveys covered only one or two fields,
typically with a field of view of $\sim50$ arcmin$^2$ due to the small
sizes of CCDs, and were strongly affected by cosmic variance
\citep[e.g.,][]{Efstathiou91, Roche93, Brainerd95, Hudon96, Woods97, McCracken00}.

More recently, wide-field cameras on intermediate-class telescopes
have led to deep wide-field photometric surveys
\citep[e.g.,][]{Neuschaefer95,Roche99,Cabanac00,McCracken01,Wilson03}.
A ground-breaking study was conducted by \cite{Postman98}, who used a
4-m telescope with a 16 arcmin field of view to cover a much wider
area by tiling together several pointings, surveying a contiguous 4 x
4 degree$^2$ field to a depth of $I_{\rm AB}=24$, reaching to $z\sim1$.
\cite{McCracken01} present a survey of 1 degree$^2$ to a
depth of $I_{\rm AB}=25.5$ over four independent fields using the UH8K
camera on CFHT, and \cite{Wilson03} use the same camera to image 1.5
degrees$^2$ over 3 separate fields to a depth of $I=24.0$.  

Other recent studies employ accurate photometric redshifts to measure
\wt \ for galaxy samples of a specific type over well-defined redshift
ranges.  \cite{Brown03} investigate the clustering of red galaxies 
in three redshift bins between $0.3<z<0.9$ in a 1.2 degree$^2$ 
field from the NOAO Deep Wide-Field Survey \citep{Jannuzi99}, while
\cite{Budavari03} use data from the Sloan Digital Sky Survey 
(SDSS, \cite{York00}) to construct a volume-limited sample with more
than 2 million galaxies between
$0.1<z<0.3$ with which to study galaxy clustering as a function of
color and spectral type. 

In this paper we
present a survey undertaken with the CFH12K camera covering a total of
5 degrees$^2$ over 5 separate fields to a depth of $I_{\rm AB}=24.0$.
There is a wide diversity of adopted redshift distributions for 
magnitude-limited samples in the literature, which can lead to different
interpretations of similar results.  
Here we use much improved $dN/dz$ distributions, relying on recent data
from the DEEP2 Galaxy Redshift Survey.  Unlike previous studies, we
require no spectral modelling in fitting for these $dN/dz$
distributions and no assumptions are made about the evolution of the
galaxy population.

An outline of the paper is as follows.
In Sections 2 and 3 we describe the observations and data reduction
process.  As a large subset of this data was used for the DEEP2 Galaxy
Redshift Survey photometry, we provide a detailed account of
 the data reduction pipeline.   
In section 4 we present galaxy counts in three photometric bands, while in
Section 5 we measure \wt \ as a function of $I_{\rm AB}$ magnitude, 
describe the $dN/dz$ distributions we use and discuss constraints 
on the evolution of galaxy clustering.  Section 6 parallels Section 5
but focuses on clustering as a function of $(R-I)$ color.  In this
section
we discuss in particular the strong clustering seen for the reddest
and bluest galaxies and discuss the most likely interpretations of
this effect.  We conclude in Section 7.

\section{Observations}

Our data were taken using the CFH12K camera on the 3.6-m
Canada-France-Hawaii telescope.  The CFH12K camera has a 12,288
$\times$ 8,192 pixel CCD mosaic array and a plate-scale of
0.207$\arcsec$ per pixel, providing a field of view of 0.70 $\times$
0.47 degrees.  We observed five separate fields on the sky, 
with one to five distinct CFHT12k pointings per field, where the area
in one pointing equals the field of view of the camera.  Our data were
taken during several observing runs from September 1999 to October
2000 and with queue-scheduled observing from April to July 2001; 
observing details are given in Table \ref{obstable}.

We integrated on each pointing for $\sim$1 hour in {\it B} and {\it R}
and $\sim2$ hours in {\it I}, broken down to individual exposures of 600
seconds.  We dithered the telescope slightly between
exposures to fill in gaps between the CCDs and to cover areas with low
charge transfer to yield well-sampled combined images.  We also took
four additional 120-second {\it I}-band astrometry images in each
pointing, with offsets of 3.5$\arcmin$ in RA and 7$\arcmin$ in 
Dec. between images.  The seeing varied from 0.69$\arcsec$ to
0.95$\arcsec$, with final seeing estimates for each pointing listed in
Table \ref{obstable}.

\section{Data Reduction}

\subsection{Initial Processing}

The data were reduced using the C-based IMCAT software package 
developed by Nick Kaiser\footnote{http://www.ifa.hawaii.edu/~kaiser/imcat/}.
After bias-subtracting each exposure, we constructed a super-flat for
each passband from the median of the science images, which were
individually normalized by the mode.  The super-flats were
adjusted by applying a multiplicative correction to each chip (aside
from one chip arbitrarily chosen as a reference) such that when
divided into the science data frames they resulted in a near constant
sky level across the entire focal plane.  We then created mask images
which masked out saturated stars, regions with bad charge transfer and
other CCD defects.  After dividing each exposure by the appropriate
super-flat we applied the mask images.  A model for the sky level
in each exposure was then obtained by estimating the local mode 
using a set of 8$\times$16 sub-images of size 256$\times$256 pixels
 and smoothly
interpolating between these points using triangular tessellation (very
discrepant values, e.g. around bright stars, were rejected). 
We then created initial catalogs of objects from each of the sky-subtracted
images, computing magnitudes using a 20-pixel radius aperture, 
which is large enough to be 
unaffected by variations in the seeing, but not so large
that systematic errors in flux from uncertainty in the background sky
level dominate. 
From these catalogs we selected stars in each exposure, for both the
astrometric and science images, selecting compact, bright objects.  
The star catalogs from the astrometry images
were then used to determine astrometric solutions for our dataset.

\subsection{Astrometry}

\begin{figure}
\centerline{\scalebox{0.5}{\includegraphics{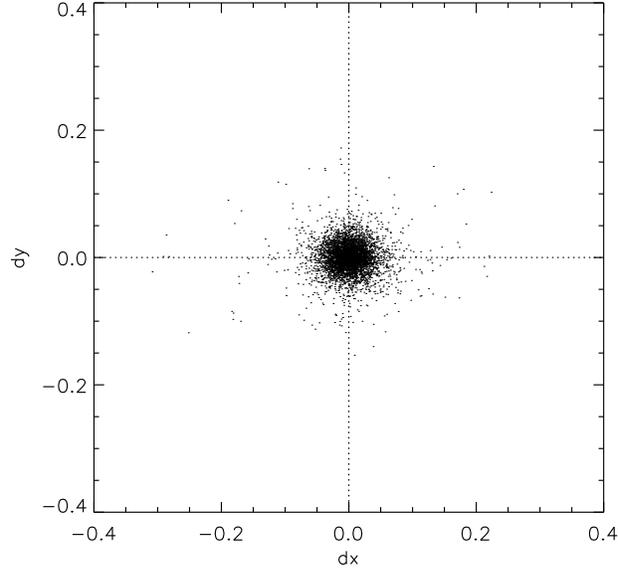}}}
\caption{Residuals in the astrometric solution, with units in pixels.
The rms is $\sim$0.025 pixels, corresponding to $\sim0.005\arcsec$.
\label{astrom}}
\end{figure}

We used the US Naval Observatory catalog (USNOA 2.0) as an absolute 
astrometric reference system.  In each pointing,
we used stars which were detected in both the USNOA catalog and in at
least one of our four astrometry images to
determine a set of true star positions from which we fit (rejecting
outliers) cubic polynomial distortion parameters for each of the 4
exposures $\times$ 12 chips. The resulting rms errors are
approximately $0''.006$, as shown in Figure \ref{astrom}, where we
plot the residuals in pixel coordinates of the stellar positions from
the best-fit solutions.  These are our relative astrometric errors;
our absolute errors are defined by the accuracy of the USNOA catalog,
which is roughly $0.5''$.

We then used these field distortion models to compute the
amplification associated with the distortion as a function of position
on the focal plane.  This can in turn be used as a check of our
astrometry solutions; we expect the amplification to be a simple
quadratic function of radius, such that gross deviations from this
form would most likely indicate systematic errors (feeding in from the
large-scale systematic errors in the USNOA catalog). The field
distortion amplifications were also needed to correct the stellar
magnitudes.  We found that the field distortion conformed well to the
expected circularly symmetric quadratic behavior, and we conclude that
systematic errors in our astrometry are very small.
After solving for the astrometric solution between the astrometry
images and the USNOA catalog, we fit for astrometric solutions between
the science exposures for each pointing and the astrometric exposures,
again using catalogs of stars selected by size and magnitude.

\subsection{Initial Photometry}

\begin{figure}
\centerline{\scalebox{0.5}{\includegraphics{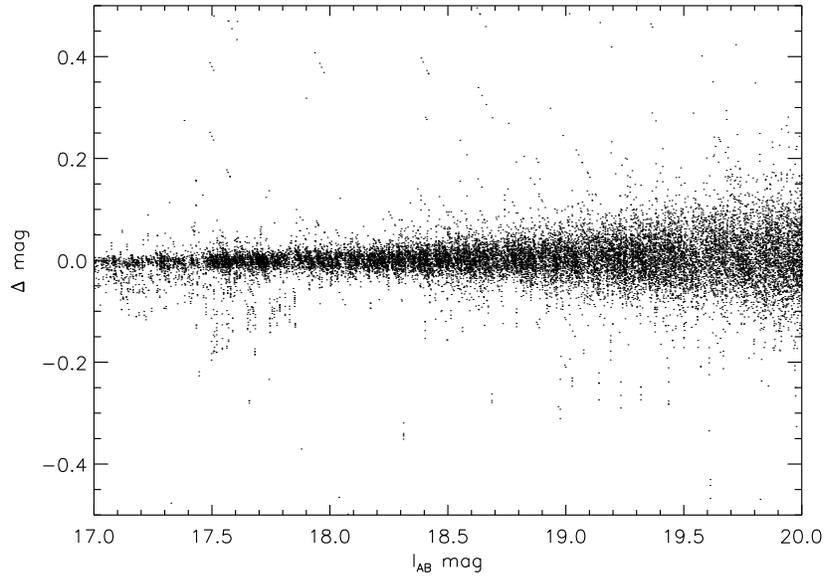}}}
\caption{Differences in magnitudes for stars detected on multiple
exposures.  The rms at bright magnitudes is $\sim$0.02.
\label{magerror}}
\end{figure}

Each science exposure yielded a catalog of $\sim800-1000$ stars.
The stellar magnitudes were corrected
for field-distortion amplification, and the catalogs of each pair
of exposures were merged using the sky coordinates provided by
the astrometric solution.  This provided a very large set of pairs of
measured stellar magnitudes, along with the chip numbers, detector coordinates
and exposure numbers.
We then fit, by least squares minimization, a model in which the
magnitude of an object measured on chip $c$ in exposure $e$ is $m_{ce}
= m + dm_c + dm_e$.  This method simultaneously measures corrections
for extinction terms in individual exposures ($dm_e$) as well as
chip-to-chip zeropoint offsets ($dm_c$).  The extinction values fit
were generally very small, typically $\sim0.005$ magnitudes,
indicating that conditions were accurately photometric, but much
larger values were found for the chip correction factors 
$dm_c$, ranging up to $\sim
0.05$mag.  This is not surprising, as the CCDs in the CFHT12K
are an inhomogeneous collection, with substantial variations in the
quantum efficiency.  The resulting error at bright magnitudes ($16.5 <
m < 18.0$) is $\langle (m_1 - m_2)^2 \rangle^{1/2} \approx 0.02$ mag,
as shown in Figure \ref{magerror}.  At fainter magnitudes, the scatter
is larger due to measurement error.
These results show that with this simple correction the magnitudes for
stars are highly reproducible exposure-to-exposure 
with an error $\sim0.02 / \sqrt{2}$.

Finally, we combined the individual science exposures for a single
pointing, deleting cosmic rays, correcting for the field distortion, 
and creating a set of $9
\times 13$ sub-images of size $1024 \times 1024$ pixels (`quilts').
For each quilt, we first determined which input images contributed;
then the appropriate sub-images were generated and combined into a set
of image planes.  These were median-combined, with rejection of
discrepant pixels.

\subsection{Catalog Construction}

Object catalogs were constructed for each quilt image using
the faint object detection algorithm described in Kaiser, Squires, \&
Broadhurst (1995).  We smoothed the $R$-band images (the deepest
obtained) with a sequence of 'Mexican hat' filters of progressively
larger size, tracked the peak trajectories through the
three-dimensional space of x, y, and Gaussian radius, and then defined
an object to be the peak of the significance (the height of the peak
divided by the rms fluctuation for that smoothing scale) along the
trajectory.  We used a filter which is a normalized Gaussian of scale
$r_g$ minus a normalized concentric Gaussian of scale 2$r_g$.  The
filter scale was varied from 0.5 to 20.0 pixels with equally spaced
logarithmic intervals $\Delta r_g=0.2r_g$; the best-fit value of $r_g$
provides an excellent estimate of the Gaussian radius of an object on
the sky, with smaller fractional scatter (for stars) than either the
half-light or Petrosian radii measured by IMCAT.  We therefore use
$r_g$ as our measure of object size throughout the remainder of this
paper.  Note that because of the limits of the grid, objects with $r_g
> 20$ pixels will be poorly measured; however, this is of little consequence 
in the magnitude ranges considered here.  All objects
with significance $\nu > 4$ were output to an initial object catalog;
we then perform photometry for each object in four different
circular apertures (of radius $3r_g$, $6r_g$, $1\arcsec$, and $2\arcsec$) 
in each band.

Unfortunately, the algorithm used to find objects can produce
false detections due to instrumental features such as diffraction
spikes, bleed trails, regions of scattered light, etc.  To remedy
this, we developed separate IDL software to identify the problematic
pixels on the projected, combined CFHT12k images; this is done separately
in $B$, $R$, and $I$.  Our algorithm begins by flagging all saturated
pixels, along with the central pixels of all partially-saturated
sources identifiable in the initial object catalog; this limit is
found empirically in each pointing as the magnitude where the measured
sizes of stars are no longer independent of their flux.  Then, all
pixels less than 3 pixels from a flagged point in both X and Y for
which the flux (after smoothing with a 3-pixel boxcar) is more than
3.25 times higher than the sky noise (also after smoothing, from the
IMCAT sky inverse variance map) are themselves flagged as bad.  This
procedure is iterated until there are no pixels with significant flux
within 3 pixels of a flagged pixel.
In what follows, all objects whose centers are within 10 pixels of any
flagged pixel (or pixel with no data, e.g. due to edges or bad
pixels in the detector) in any band are rejected from analysis; this
radius corresponds to $\sim4 r_g$ for a typical object, and thus is
larger than the final apertures used.  The bad pixel masks produced
by this procedure are also employed in masking the random catalogs used 
for correlation analyses (see Section 5.1).

We found by visual inspection of the resulting masked catalog that
occasionally there were spurious or incorrect entries, either due to
overlapping objects, where substructure within a large, extended
object is detected as a small separate object within its bounds, or
``over-merged'' pairs or triplets of galaxies which were identified
incorrectly as one extended object.  To rectify these situations, we
applied two algorithms, one to delete overlapping objects and one to
split merged objects.  For the overlapping objects, we first define a
'small' radius, $r_s=0.85 r_{seeing}$, and then flag as spurious any
objects with $r_g<r_s$ and which are inside 3$r_g$ of a bright,
extended object or that have $r_g<1.5*r_s$ and are within 3$r_g$ of a
bright, extended source and have peak counts more than 2 sky sigma
below the peak of the nearby background (i.e. lie on a very strong
gradient background).  For all tests of object overlap, we use
ellipsoids with $r_g/3r_g$ as the semimajor axis, which matched visual
morphologies well.  As a part of this routine, we also eliminate from
the catalog identical overlapping objects identified on multiple quilts.

For the over-merged objects, we first flag all low surface-brightness
(LSB), elongated objects, with $b/a \lesssim0.5$, as many of these are
multiples which have been incorrectly designated as a single object.
Around each flagged object we smooth the image quilts with a Gaussian
kernel of $\sigma=0.67 \sigma_{seeing}$ and find all peaks within 
a 2-pixel radius circle which are
greater than five times the sigma of the smoothed sky (i.e., the
highest surface brightness regions present).  We then centroid on
these and measure their FWHM on the image quilt.  We evaluate the
significance of the original object by determining whether, after
masking the candidate sub-objects, a fit to an elliptical-Gaussian
model has an amplitude greater than $5\sigma$ above 0 (where here
$\sigma$ is the uncertainty in this Gaussian-weighted amplitude).  If
not, we give the original object a quality flag such that it will 
be rejected from further use.  If the significance of
detection for the original object is $>30$,
then we consider all the high surface brightness (HSB) regions to be
merely bright sub-regions of an original object, and reject them.
Otherwise, for all the HSB regions, we evaluate: 1) whether they are
resolvable from other HSB regions identified in this process -- in
case of such conflicts, we keep only the brightest peak; 
2) if the HSB object is indistinguishable from a known
object in the catalog -- i.e. we require its distance from each known
catalog object be: a) $>1.25 \sigma_{seeing}$ from the catalog object, such 
that separation of the peak is resolved, or b) 0.67 times the catalog
object profile $\sigma$ from central object, so that we are not likely to
be seeing substructure in the peak; and 3) whether the identified
object in fact appears diffuse -- i.e., has FWHM$>2.5$ times the
seeing FWHM, and not identified with a catalog object.  This throws
out false diffuse detections due to low-level flat-field or sky-subtraction
issues that are not the over-merged objects that we are
trying to eliminate.  All of the criteria and limits used were tuned
empirically based on the actual over-merging of objects seen.  
We found that $\sim3$\% of the objects were 
``over-merged'' and were thus split into separate objects, on which
we performed photometry in all three bands.   
As colors are computed in $1\arcsec$ apertures, 
biases in the colors due to blending of sources are minimized.

\subsection{Star-Galaxy Separation}

The magnitude range of our data spans a regime where stars are a
non-negligible contaminant for a galaxy catalog.  Unfortunately, size
and magnitude alone are not always sufficient to distinguish stars
from galaxies, especially at fainter magnitudes where
small intrinsic sizes and uncertainties in size measurements cause
some galaxies to have measured sizes comparable to those of stars.  We
have therefore developed a set of algorithms which distinguish stars
from galaxies in a probabilistic fashion, using their sizes,
magnitudes, and $(B-R)$, $(R-I)$ colors.  We exploit the fact that
stars and compact galaxies have very different magnitude
distributions, with stars dominating at bright magnitudes and galaxies
at faint magnitudes, and that stars fall within a tight locus in
color-color space.  Objects which are clearly extended, with 
$r_g>0.05$ above the upper limit of the locus of stars in
the size-magnitude plane, are deemed galaxies, while compact objects
are given a probability, in a Bayesian sense, of being a galaxy given
their magnitude $m$ and color vector ${\bf c}$:
\begin{equation}
p_{gal}(m,{\bf c})=\frac{p({\bf c}|gal) p(gal|m)}{p({\bf c}|gal)
p(gal|m) + p({\bf c}|star) p(star|m)}.
\label{pgal}
\end{equation}

The magnitude distribution of compact objects is essentially linear at
the bright end ($R\lesssim21$) and rises exponentially at fainter
magnitudes.  We assume, quite reasonably, that the linear 
component is due exclusively to stars, while galaxies cause the 
exponential increase at the faint end.  In each pointing, we perform
a non-linear least-squares fit to the magnitude distribution function
$n(m)=a_0 e^{a_1 m}+a_2+a_3 m$, where $m=m_R-22$.  We then define the
probability of a compact object being a galaxy, based on its
magnitude, as $p(gal|m)=(a_0 e^{a_1 m})/n(m)$, with
$p(star|m)=1-p(gal|m)$. In practice we limit the range for these
priors, confining them to the range 5\%-95\%, in order to restrain the
influence of any errors in our assumptions.  It then remains to
determine the color distributions of stars and galaxies.

It is first necessary to bring all pointings to a
common color system.  We do an initial, rough calibration using
observed number counts; in the absence of cosmic variance, the number
of galaxies within a given magnitude range should be the same in each
pointing and bandpass.  We can refine the zeropoints of our color
system by forcing the
locus of stars, chosen for their compact size and bright magnitudes,
for the first iteration of this procedure, in color-color space to
match amongst all pointings.
This locus is well-approximated in our bands by two intersecting line
segments of fixed slope; therefore, by fitting the color distribution of
stars in each pointing to that functional form, via least-squares 
minimization, we
may derive the offset in both $(B-R)$ and $(R-I)$ colors required to
make the stellar locus match our reference pointing.  We chose 
pointing 2 in field 2330 (see Table \ref{obstable}) to be our reference
pointing, for reasons described below.  This
procedure indicates that the color zeropoints derived from number
counts are consistent from pointing to pointing to within 
0.03 mag rms in $(R-I)$ and 0.05 mag rms in $(B-R)$.
 By using the stellar locus, although we do not necessarily know
the correct absolute zeropoint in any pointing in any band, we can
bring all pointings to a common color system in a robust way.  We
iteratively repeat this stellar locus calibration as the color
distributions of stars and galaxies are refined, using the
probabilistic star-galaxy separation defined above after the first
color distributions are constructed; excellent convergence occurs within two
iterations.
In addition, we use the variations in the stellar locus within
subregions of each individual pointing to derive a smooth correction
for residual scattered light and/or flat field errors, applied using a
minimum-curvature surface interpolation.  

\begin{figure}
\centerline{\scalebox{0.6}{\rotatebox{90}{\includegraphics{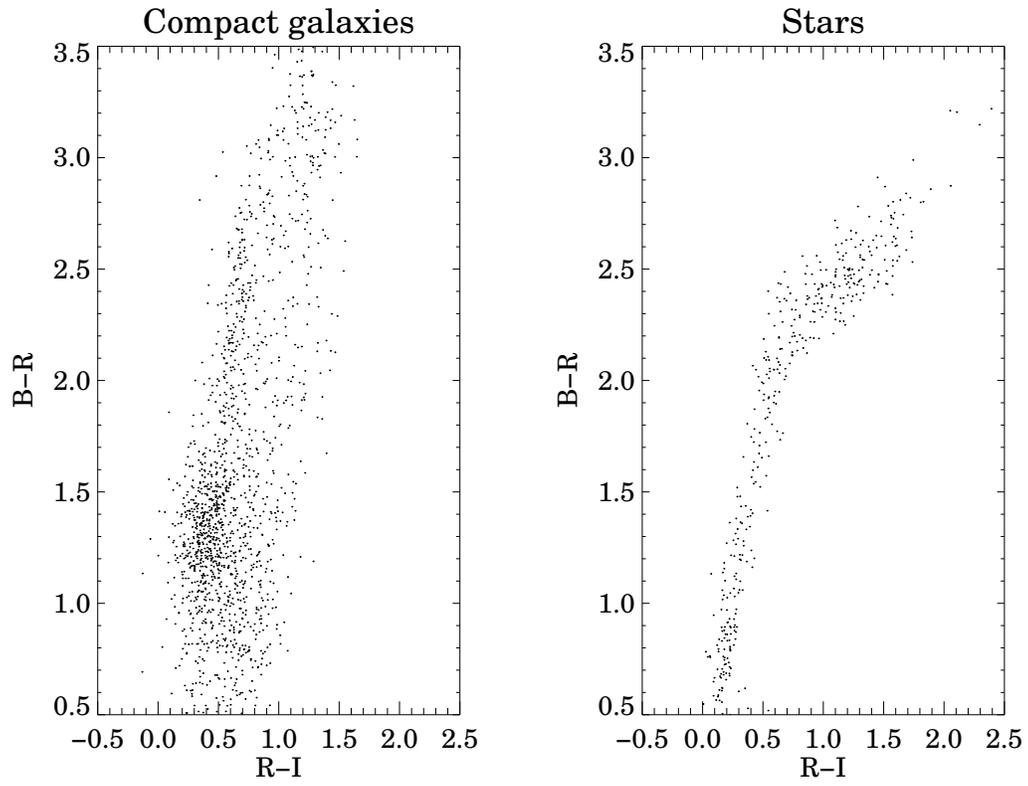}}}}
\caption{Color-color distribution for compact galaxies and stars in one
pointing; these distributions are used to
define the probability that a given object is a galaxy.
\label{pgalfig}}
\end{figure}

After bringing all pointings to a common color system, we select
objects within a pointing which are: 
1) almost certainly stars (i.e., that have $r_g$ within
the stellar range, and either $19<m_R<21.25$, or $m_R < 24$,
$(R-I)>1.45$, and color within $\pm 0.2$ mag of the expected stellar
locus -- this latter requirement is necessary to include M stars),
2) bright compact galaxies ($20.5<m_R<23.5$, $r_g$ larger than the
stellar range but less than $\sqrt{(1.5 pix)^2+r_{seeing}^2}$), or
3) faint compact galaxies ($m_R>23.5$ and $r_g$ as above).  
Plots of the positions for compact galaxies and stars 
in $(B-R)$ and $(R-I)$ space for data in one pointing are shown in Figure \ref{pgalfig}.  

\begin{figure}
\centerline{\scalebox{0.4}{\includegraphics{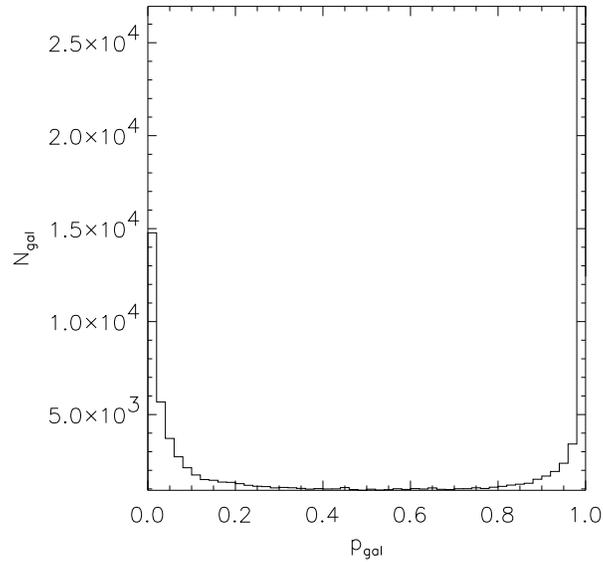}}}
\caption{Probability distribution that a compact object is a galaxy for
data in all pointings.  The distribution is quite bi-modal; most 
compact objects are clearly classified as stars or galaxies. 
\label{pgaldistfig}}
\end{figure}

We then create a two-dimensional color-color histogram for each of
these samples, which we smooth by a two-dimensional Gaussian kernel
with a $\sigma=0.05$ mag to take into account the effects
of magnitude errors in applying these distributions to catalog
objects.  We delete outliers, redo the smoothing, and normalize the
distributions to have a maximum probability of one.  This results in a
probability distribution for the color of stars directly, though for
galaxies we have to take into account the magnitude-dependence of
their distribution in color-color space, so as not to be dominated by
the more numerous, faint objects.  We therefore define $p({\bf
c}|gal)=w(m_R) p({\bf c}|bright) +(1-w(m_R))p({\bf c}|faint)$, where
$bright$ and $faint$ denote the two galaxy samples we defined above
and $w=1$ for $m_R<23$, $w=m_R-23$ for $23<m_R<24$ and $w=0$ for
$m_R>24$.  We then use equation \ref{pgaldistfig} to determine the
probability for each compact object that it is a galaxy. 

We emphasize that we have computed color-color distributions for
the bright and faint compact objects separately; we are not assuming
that galaxy colors and magnitudes are completely independent.  
It is primarily important in our Bayesian method that the color 
distribution of galaxies -- our prior -- fills a different space than 
the stellar color distribution.  And as galaxies of a given redshift 
and restframe color have substantial breadth to their luminosity function, 
the covariance between color and magnitude is dilute.  
A full treatment of the three-dimensional distribution of magnitude 
and color would greatly complicate implementation and would yield noisy 
results at bright magnitudes where we have fewer galaxies to build a 
distribution from, but also where star-galaxy separation is most important.
The resulting strong separation between stars and galaxies that we find using
 our probabilistic method shows that we are not sensitive to such details.

We find that 68\% of the objects are extended, while 2\% do not have
sufficient information to determine $p_{gal}$, either because they lack
data in one or more passbands or are brighter than the saturation limit.
Figure \ref{pgal} shows the $p_{gal}$ distribution for the {\it compact} 
objects, which is extremely bi-modal, 
with 25\% of the objects having 0\% probability of being a galaxy and 38\%
having 100\% probability of being a galaxy.  
The few objects with intermediate values of $p_{gal}$ tend
to lie near these extremes, and so in the analyses of angular 
correlations that follow we include only those objects with a 
$>20$\% probability of being a galaxy, though it makes very little difference
what exact cutoff is used.  Effects on the angular clustering measurements
due to stellar contamination are discussed in section 5.2.

\subsection{Final Photometry}

A large portion of the photometry presented here is used by the DEEP2 Galaxy 
Redshift
Survey \cite{Davis02}, and as a courtesy we define the photometric 
calibrations such that they are optimal for that survey.  
We therefore define a set of magnitudes in the AB system, with a 
total, $R_{\rm AB}$-band magnitude and $(B-R)$, $(R-I)$ colors for each
object. We adopt as our measure of total magnitude the $R$
magnitude measured in a circular aperture of radius $3r_g$, 
unless $3r_g$ is less than $1\arcsec$ (due to noise in its measurement), 
in which case we use the magnitude within a $1\arcsec$ radius.
We found that the 
magnitude within $6r_g$ as measured by IMCAT is often contaminated by
nearby objects included within the aperture or problematic as the
aperture goes beyond the buffer region of each quilt for objects near
their edges, while the $3r_g$ magnitude proved more robust. 

Conventionally, we would then use the $B$ and $I$ magnitudes measured
within this same aperture to define our system; however, due to
differences in seeing/aperture corrections between bands and amongst
the CFHT12k pointings, it would then be impossible to make the colors of
extended objects (i.e. galaxies) agree pointing-to-pointing even if
the colors of the stellar locus match; this effect can reach $\sim
0.05$ mag in each band.  This effect is much smaller for colors
derived from the IMCAT $1\arcsec$ aperture magnitudes, $\lesssim
0.02-0.03$ mag in both $(B-R)$ and $(R-I)$, based on cross-correlating the
color locus of extended objects from pointing to pointing.
Therefore, we adopt as the basis of our magnitude system the
quantities we can measure and calibrate most robustly:  the
``total'' $R_{\rm AB}$ magnitude, measured in the $3r_g$ aperture, and the two
colors ($B-R$) and ($R-I$), each measured within $1\arcsec$ 
apertures.  In the remainder of this paper, $B$ will be used to
represent $R_{total}+(B-R)_{1\arcsec}$, and $I$ will represent
$R_{total}-(R-I)_{1\arcsec}$.  Note that these should be distinct from 
conventional magnitudes only in the case of resolved color 
gradients within the observed galaxies. 

We calibrate our $R_{\rm AB}$-band magnitude and $(B-R)$, $(R-I)$ colors 
through two separate processes.  First, we calibrate $B$, $R$, 
and $I$ to an AB magnitude system in the native CFHT12k passbands 
using SDSS 
photometry for stars in our reference pointing (pointing
2 in field 2330 [see Table \ref{obstable}]).
This pointing was chosen to define our color system as it
yielded the cleanest comparison to SDSS, due to the quality both
of the CFHT12k observing conditions and the SDSS photometry (it lies
on the celestial equator and has been observed by SDSS repeatedly).
The accuracy of this calibration is $\sim0.02-0.03$ mag (worst in
$B$), dominated by errors in the SDSS calibration.  Through
iteratively fitting the stellar locus, bringing all pointings to a
common color system, determining the overall color 
distributions of stars and galaxies, and reapplying our star-galaxy
separation algorithm, we not only produce a stable star-galaxy
separation but also bring all pointings to the same $(B-R)_{1s}$,
$(R-I)_{1s}$ color system to within $\sim0.01$ mag in each color; the
zeropoints of that color system are then set by the SDSS calibration
of our reference pointing.  After our correction for spatial variations in the
color system using variations in the stellar locus, residual
variations in the $(B-R)$ and $(R-I)$ zeropoints within each pointing
are estimated to be 0.04 and 0.03 mag, respectively; these
variations should primarily be relevant on scales of $\sim4\arcmin-8\arcmin$.

Second, after this calibration of the color system has been performed, we 
calibrate the $R$-band zeropoint in each pointing.
In the four pointings for which no information from SDSS is available
(and which are not used in the DEEP2 Galaxy Redshift Survey),
we determine zeropoints by forcing the number counts to match those from our
other fields; based on pointings with SDSS calibration, we find that
this yields a zeropoint with errors of 0.10 mag, due to cosmic
variance (errors from this technique in $(B-R)$ or $(R-I)$ are smaller
as the cosmic variance is strongly covariant between bands).  We have
tested several of the below analyses with and without the inclusion of
these less-calibrated pointings, and all differences were well within the
quoted errors.  In the fields with SDSS calibration,
however, by comparing $B$, $R$, and $I$ to the SDSS magnitudes for
stars, we get three {\it independent} estimates of the $R$ zeropoint
(as we have tied the three magnitudes together in our photometry by
making the stellar loci match).  In general, the zeropoints estimated
from $R$ and $I$ agree to within 0.02 mag rms, while these agreed with
the zeropoint from $B$ to $\sim0.04$; this matches well the expected
variation in SDSS photometric zeropoints.  We use the average of the
zeropoint offsets from $R$ and $I$ to correct our photometry wherever
SDSS photometry is available.  Based on comparisons to SDSS, residual
variations of the $R$ zeropoint within each pointing, which we cannot
correct for with either the stellar locus or number counts, appear to
be $\lesssim0.02$ mag rms.
The final uncertainty in the absolute calibration includes both 
a random error due to the finite number of stars with well-measured
magnitudes in both datasets and
a systematic error due to 
uncertainties in the AB magnitude calibration of the SDSS photometry.

\section{Galaxy Counts}

\begin{figure}
\centerline{\scalebox{0.8}{\includegraphics{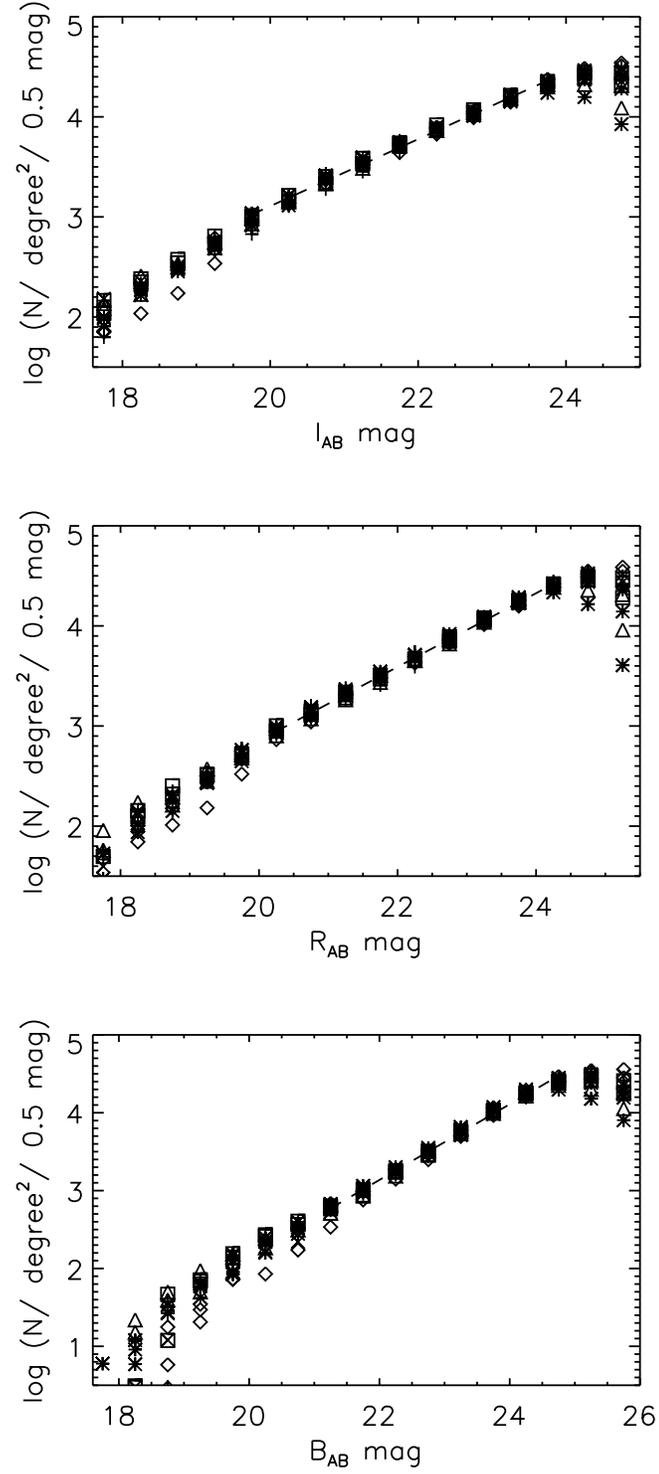}}}
\caption{Galaxy counts for $B_{\rm AB}$, $R_{\rm AB}$, and $I_{\rm AB}$ 
bands.  The dotted lines are linear fits discussed in the text (see Section 4).
\label{galcounts}}
\end{figure}

We first present galaxy number counts for our data in order to test
the uniformity and depth of the sample.  Figure \ref{galcounts} shows
histograms of $B_{\rm AB}$, $R_{\rm AB}$ and $I_{\rm AB}$ 
galaxy counts for all 15
pointings, in units of log number of counts per degree$^2$ per 0.5
magnitude.  At the bright end we see
variations in the counts due to cosmic variance, while at the faint
end the varying depth of the data in each pointing leads to
differences in the counts past our completeness limit.
 Our data is complete to $\sim25.25$ in $B_{\rm AB}$,
$\sim24.75$ in $R_{\rm AB}$, and $\sim24.25$ in $I_{\rm AB}$ in all
pointings except field 1416, pointing 2, which is our shallowest
pointing and the low data point at each of the completeness
limits in Figure \ref{galcounts}.  We fit for the slope in the galaxy
counts, $\delta\equiv d \ log N/dm $, in each band over the magnitude 
range in which the data is
consistent from pointing to pointing.   The {\it
B}-band galaxy counts have a slope of $\delta=0.49 \pm0.01$ in the
magnitude range $B_{\rm AB}=21.25-24.75$ mag.  This fit is shown in the
bottom of Figure \ref{galcounts} as the dotted line and agrees well
with \cite{Brunner99} ($\delta=0.51$, $B_{\it AB}\sim20-24$),
\cite{McCracken01} ($\delta=0.47 \pm0.02$, $B_{\it AB}=20-24$) and
\cite{McCracken03} ($\delta=0.45 \pm0.01$, $B_{\it AB}=20-24$).  Our
$R_{\rm AB}$ galaxy counts have a slope of $\delta=0.37 \pm0.01$ in
the magnitude range $R_{\rm AB}=20.25-24.25$ mag, in agreement with
\cite{Hogg97}, \cite{Brunner99}, \cite{Metcalfe01} and 
\cite{McCracken03} who find
slopes of $\delta=0.33, 0.34, 0.37$, and $0.37 \pm0.03$, respectively.
Our $I_{\rm AB}$ counts have a slope of $\delta=0.33 \pm0.01$ in the
magnitude range $I_{\rm AB}=19.75-23.75$ mag, which also agrees well with
other published results in similar magnitude ranges, all of which
report a slope between $0.31-0.35$ (\cite{Brunner99},
\cite{Arnouts99}, \cite{Metcalfe01}, \cite{McCracken01},
\cite{Wilson03}, \cite{McCracken03}).

\section{Galaxy Clustering as a Function of Magnitude}

In this section we present measurements of \wt \ as a function of
$I_{\rm AB}$ magnitude and use redshift distributions of galaxies
from the early DEEP2 Galaxy Redshift Survey data to constrain models of
clustering evolution.  Section 6 presents a parallel
study of galaxy clustering as a function of $(R-I)$ color.

\subsection{The Angular Two-point Correlation Function}

The projected angular two-point correlation function \wt \ is a
measure of the probability above Poisson of finding two galaxies with
a separation $\theta$, defined as
\begin{equation}
dP = N [1+\omega(\theta)] d\Omega
\end{equation}
where $N$ is the mean number of galaxies per steradian and d$\Omega$
is the solid angle of a second galaxy at a separation $\theta$ from a
randomly chosen galaxy.
To measure \wt \ one must first construct a catalog with a random
spatial distribution and uniform density of points with the same
selection criteria as the data, to serve as an unclustered
distribution with which to compare the data.  For each of our 15
pointings we created a random catalog with 200,000 points (17 times
larger than the largest data sample we use here) with the same sky 
coverage as our data, for which we then applied the same masking 
that was used to mask bad regions of the data due to saturated stars 
and CCD defects.

We measure \wt \ using the \cite{Landy93} estimator,
\begin{equation}
\omega(\theta)=\frac{1}{RR}\left[DD \left(\frac{n_R}{n_D}
\right)^2-2DR\left(\frac{n_R}{n_D} \right)+RR\right],
\label{dd}
\end{equation}
where $DD$, $DR$, and $RR$ are pair counts of galaxies in the
data-data, data-random, and random-random catalogs with separation
$\theta + \delta \theta$, and $n_D$ and $n_R$ are the mean number
densities of galaxies in the data and random catalogs.  This estimator
has shown to be relatively insensitive to the size of the random
catalog and handles edge corrections well \citep{Kerscher00}.  We
measure \wt \ in logarithmic bins of 0.2 in log($\theta$) with
$\theta$ measured in degrees.

Measurements of \wt \ are known to be low by an additive factor known
as the ``integral constraint'' (IC, see e.g. Peebles 1980\nocite{Peebles80}) 
which results from using the data
sample itself to estimate the mean galaxy density.  This correction
becomes important on scales comparable to the survey size, here equal
to 0.5 degrees, the short axis of a single CFHT12k pointing.  We therefore
constrain our fits to a maximum $\theta$ of 0.05 degrees, where the IC
is negligible.

\subsection{Galaxy Clustering as a Function of $I_{\rm AB}$ Magnitude}

\begin{figure}
\centerline{\scalebox{.9}{\rotatebox{90}{\includegraphics{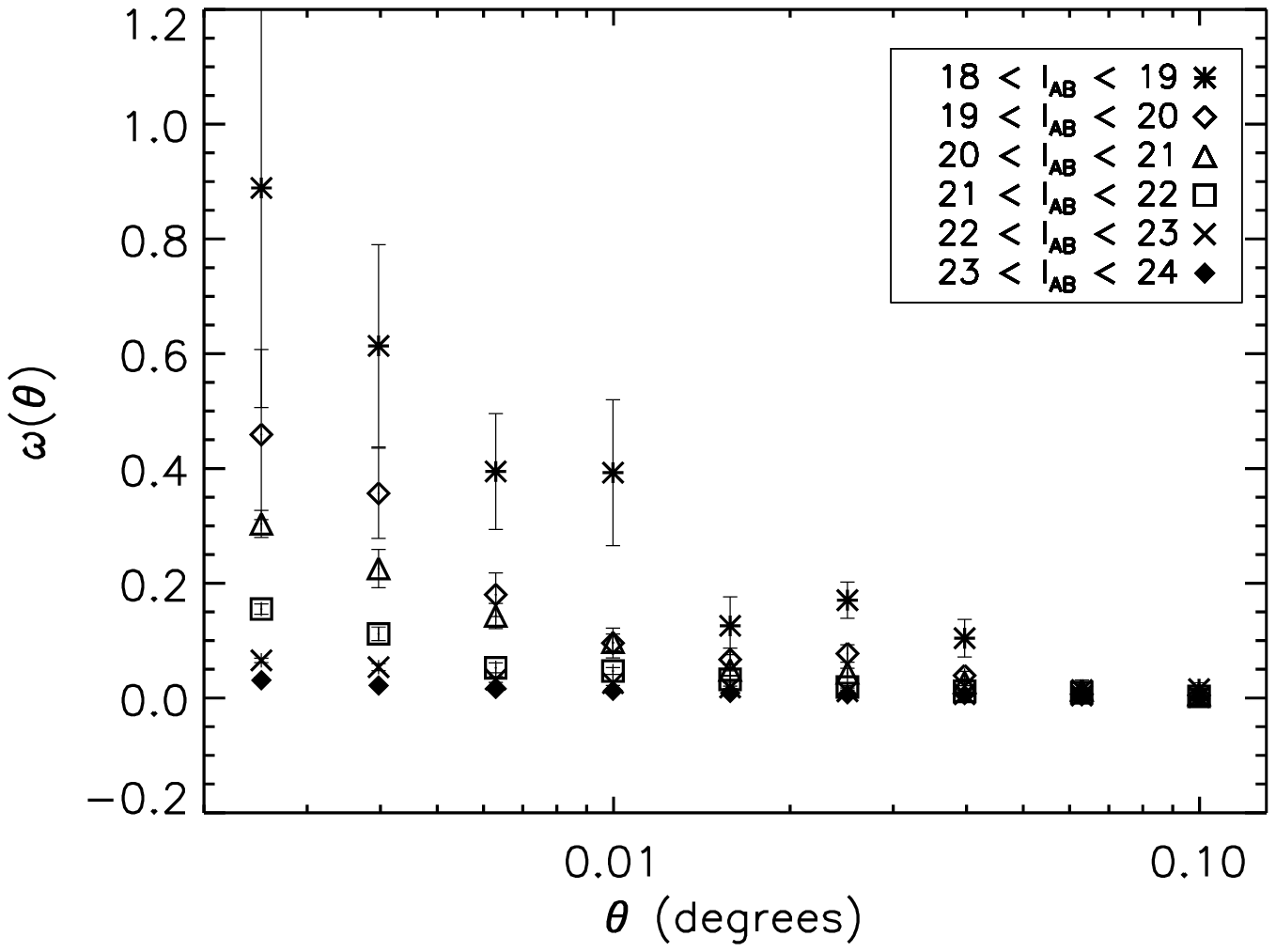}}}}
\caption{The angular correlation function, \wt, measured for 
independent $I_{\rm AB}$ magnitude bins, shown on scales 
$\sim7\arcsec-6\arcmin$.
 \label{fitmagsnolog}}
\end{figure}

\begin{figure}
\centerline{\scalebox{.9}{\includegraphics{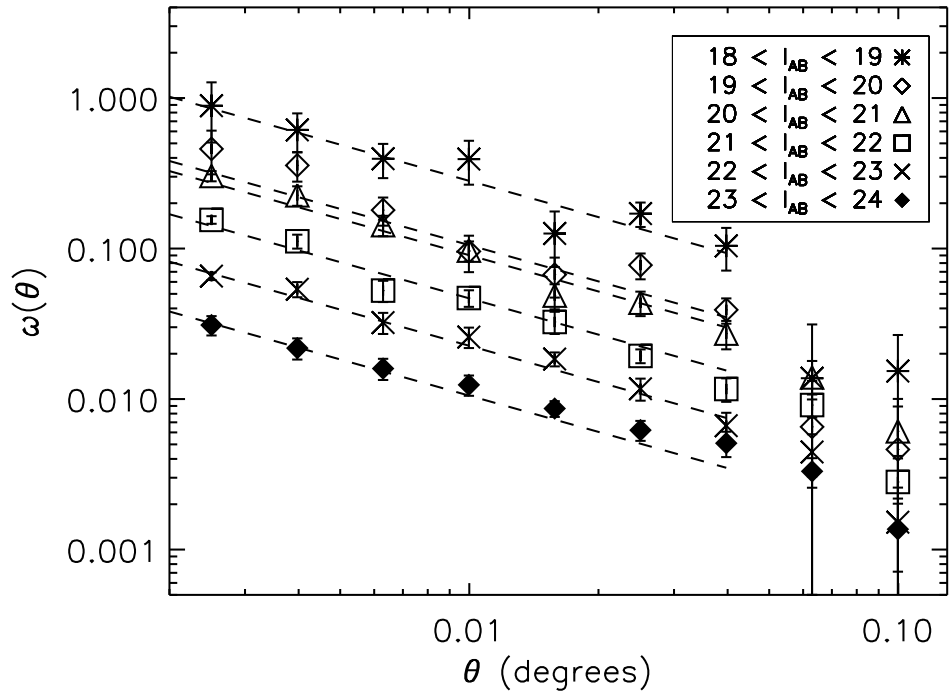}}}
\caption{The angular correlation function, \wt, plotted on a log scale 
for different $I_{\rm AB}$ magnitude ranges. 
Power law fits for \wt$=A_w \theta^{\delta}$ on scales 
$\sim7\arcsec-2\arcmin$ are shown
as dashed lines, assuming a constant slope of $\delta=-0.8$. 
 \label{fitmags}}
\end{figure}

Figures \ref{fitmagsnolog} and \ref{fitmags} present  
\wt \ measured for independent magnitude
bins in the range $I_{\rm AB}=18-24$ on scales of $\sim7\arcsec-
6\arcmin$. The amplitude 
of the clustering is seen to smoothly decrease with magnitude, with 
fainter magnitude bins showing less projected clustering. 
In Figure \ref{fitmagsnolog} one can see that \wt \
approaches zero on large scales, while Figure \ref{fitmags} shows clearly
the effects of the IC; on scales larger than $\sim3\arcmin$ the
amplitude of \wt \ begins to fall below the power-law form seen on smaller
scales.  In each of these plots we have measured the mean value of \wt \ in
each $\theta$ bin.  
The error bars on \wt \ are calculated from the 
standard error across our 15 independent pointings and should be a 
robust estimate of the error due to cosmic variance.  
We find that \wt \ measured by
first summing the DD, DR, and RR counts over all the pointings, and then
using Equation (\ref{dd}) to estimate the clustering amplitude, is 
consistent with the mean values shown here.

For each magnitude range we fit a power-law form,  
\wt$=A_{\omega}\theta^{\delta}$, on scales $\sim7\arcsec-3\arcmin$ 
and solve for the slope and amplitude using
a least-squares fitting algorithm where we minimize $\chi$-squared,
taking into account the covariance between bins as determined from
bootstrap resampling.  We first
assume a slope of $\delta=-0.8$ to compare with other published
results.  Table \ref{magtable} lists the amplitude of \wt \ measured
at 1$\arcmin$, \Aw$(1\arcmin)$, in each magnitude bin for a slope of
$\delta=-0.8$, along with an error from the least-squares fitting
and the median number of galaxies within a single pointing in that
magnitude range, equal to the median sample size.  We find that the
errors estimated from bootstrap resampling are nearly identical to
those calculated above.  In Figure
\ref{fitmags} these fixed-slope fits are plotted as dashed lines.  As
can be seen in this figure, a slope of $\delta=-0.8$ is generally
consistent with the data for the magnitude ranges and scales we study here.

\begin{figure}
\centerline{\scalebox{.6}{\rotatebox{90}{\includegraphics{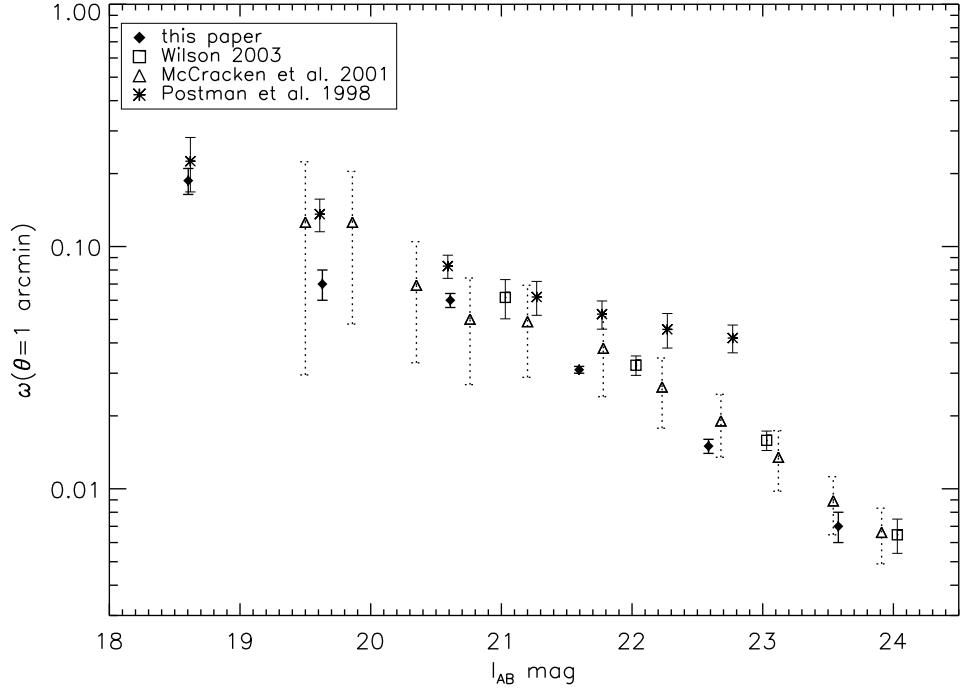}}}}
\caption{The amplitude of \wt \ measured at $1\arcmin$ shown 
as a function of the median $I_{\rm AB}$ magnitude, assuming a slope of 
$\delta=-0.8$.  Results from
\cite{McCracken01} and \cite{Wilson03} are plotted as well, with dotted error
bars for clarity. \label{wat1arcmin}}
\end{figure}

In Figure \ref{wat1arcmin} we plot the amplitude of \wt \ at
1$\arcmin$, \Aw$(1\arcmin)$, as a function of median $I_{\rm AB}$ 
magnitude for an
assumed slope of $\delta=-0.8$.  Our values agree well with the recent
results of \cite{McCracken01} (who do not use independent magnitude
bins, but always keep a bright limit of $I_{\rm AB}=18.5$ and vary the
faint limit of their magnitude range) and with \cite{Wilson03}, though
our data points are slightly lower (see
\cite{Wilson03} for a figure comparing their data with other results
in the literature).  However, our results do not agree well with
\cite{Brainerd95}, \cite{Woods97}, or \cite{Postman98}, who all find
significantly higher amplitudes at faint magnitudes, $I_{\rm AB}>22$.
The surveys of both \cite{Brainerd95} and \cite{Woods97} are quite
small, covering less than 150 arcmin$^2$ each, and are therefore 
subject to significant cosmic variance. 
\cite{Postman98} cover a very wide area (16 degrees$^2$) separated 
over 256 contiguous individual pointings.  We find that 
their values of \wt \ are a factor of two higher than our measurements 
at $I_{\rm AB}\sim22$.
It is unlikely that this large of a discrepancy could be caused 
primarily by errors on their estimates of 
the IC, which would only affect larger scales than we are considering here, 
or by errors in stellar contamination estimates, which would
have to be unreasonably large at these magnitudes.  A much more likely 
explanation is that these differences may be the result of calibration 
errors in their photometry, as any zero-point differences across their  
pointings would lead to a systematic increase in the measured clustering 
on all scales, as pairs across pointings are used on even the smallest 
scales.  This would also have a greater effect at fainter magnitudes, 
where the relative photometric error is larger. 
Here we measure \wt \ only within individual
pointings and do not attempt to compute \wt \ across pointings.
The continued decrease in clustering which we
find for fainter galaxies is also seen by \cite{McCracken01} and
\cite{Wilson03}.

\begin{figure}
\centerline{\scalebox{.6}{\rotatebox{90}{\includegraphics{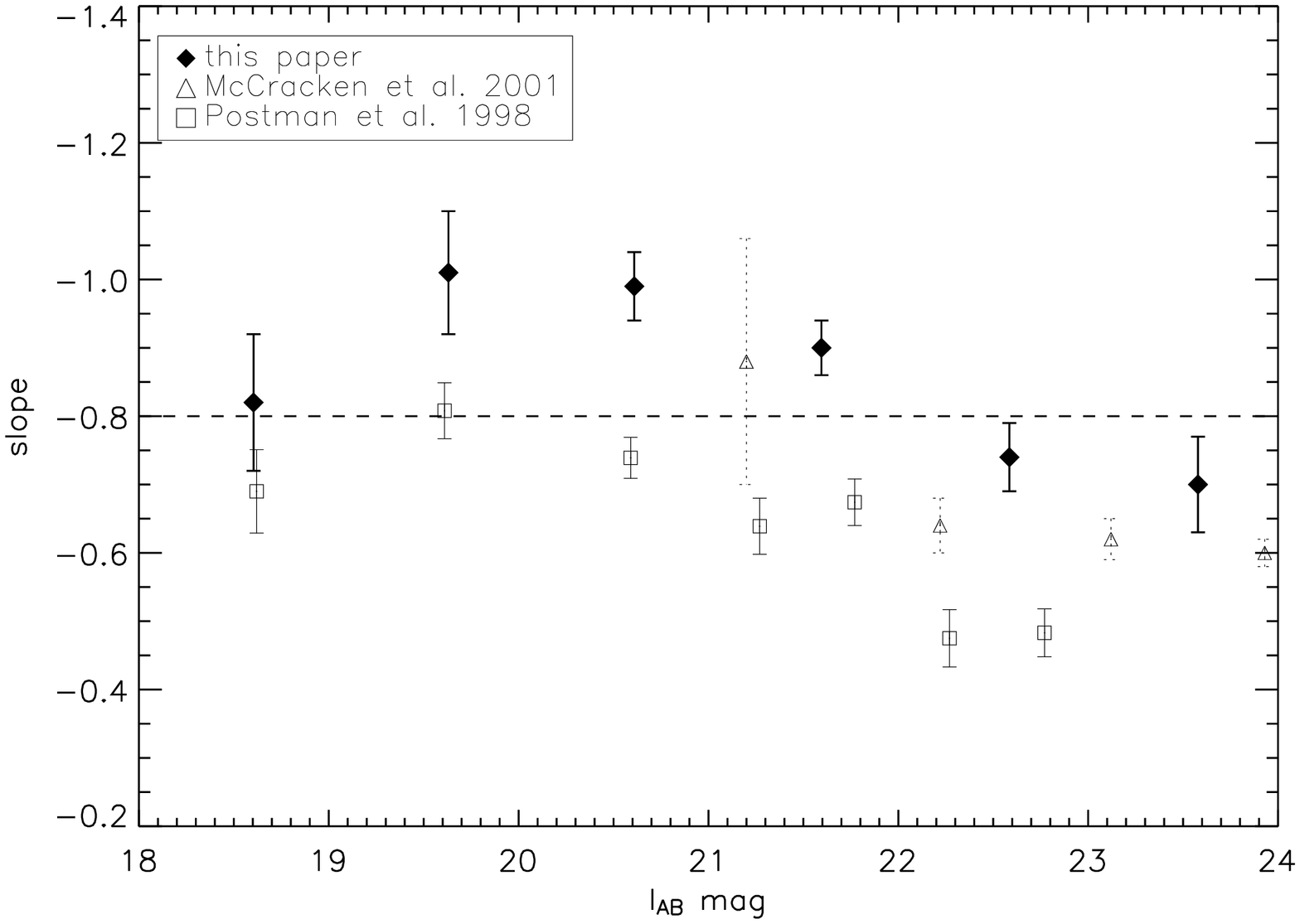}}}}
\caption{The best-fit slope, $\delta$, of \wt \ as a function of median 
$I_{\rm AB}$ magnitude.  
The slopes are fit on scales of $\sim7\arcsec-3\arcmin$. Results 
from \cite{McCracken01} and \cite{Postman98} are shown for comparison. 
\label{slope}}
\end{figure}

As can be seen in Figure \ref{fitmags}, the canonical slope of
$\delta=-0.8$ does not the fit the data exactly at all magnitudes, and
so we next relax the assumption of a fixed slope and measure both the
slope and amplitude of \wt \ for each magnitude bin on scales of 
$\sim7\arcsec-3\arcmin$.  The resulting
slope as a function of median $I_{\rm AB}$ magnitude is shown in
Figure \ref{slope} and listed in Table \ref{magtable}, along with the
amplitude, \Aw$(1\arcmin)$.  The slope
varies between $\delta\sim-0.7 - -1.0$, with a maximum
between $I_{\rm AB}\sim19-21$ and shallower values at fainter
magnitudes.  As shown in Figure \ref{slope}, our values are consistent 
with those of \cite{McCracken01},
and the general trend we find in how the slope varies with 
magnitude is consistent 
with \cite{Postman98}, though they measure shallower slopes at all
magnitudes.  \cite{Wilson03} report
that a slope of $\delta=-0.8$ fits their data well for the magnitude
range $I_{\rm AB}=20-24$.

A decrease in the slope of \wt \ at fainter magnitudes implies evolution
in the slope of \xir, such that the slope is shallower at higher redshift.
Semi-analytic models predict a decrease in the slope of \xir \ from $\sim1.85$
at $z\sim0$ to $\sim1.6$ at $z\sim1$ \citep{Kauffmann99b}, and measurements 
of \xir \ in early data from the DEEP2 Galaxy Redshift Survey have a slope of 
$\gamma=-1.66 \pm0.12$ \citep{Coil03xisp}.  These results presented here are
consistent with a similar decease in the slope at $z\sim1$ (see 
Section 5.4 for a discussion of the redshift distribution as a function of
magnitude).

\subsection{Modelling \wt}

As the two-dimensional galaxy clustering seen in the plane of the sky
is a projection of the three-dimensional clustering, \wt \ is directly
related to its three-dimensional analog $\xi(r)$.  For a given
$\xi(r)$, one can predict the amplitude and slope of \wt \ using
Limber's equation, effectively integrating $\xi(r)$ along the redshift
direction (e.g. Peebles 1980)\nocite{Peebles80}. 
 If one assumes $\xi(r)$ (and thus \wt)
to be a power-law over the redshift range of interest, such that
\begin{equation}
\xi(r,z)=\left[\frac{r_0(z)}{r} \right]^\gamma,
\end{equation}
then
\begin{equation}
w(\theta)=\sqrt{\pi} \frac{\Gamma[(\gamma-1)/2]}{\Gamma(\gamma
/2)}\frac{A}{\theta^{\gamma-1}},
\label{wthetaeqn}
\end{equation}
where $\Gamma$ is the usual gamma function.  
Here we use the convention of $r$ being a physical length, and $x$ being a
comoving length, such that $r=ax$, where $a$ is the scale factor. 
We therefore quote the comoving 
scale-length of clustering as a function of redshift as $x_0(z)$ and 
the local value as $x_0(0)$ (equal to $r_0(0)$). 
The amplitude factor, $A$, is given by
\begin{equation}
A=\frac{\int_{0}^{\infty} x_0^\gamma(z)
g(z)\left(\frac{dN}{dz}\right)^2
dz}{\left[\int_{0}^{\infty}\left(\frac{dN}{dz} \right) dz \right]^2}
\label{Aeqn}
\end{equation}
where $dN/dz$ is the number of
galaxies per unit redshift interval and $g(z)$ depends on $\gamma$,
and the cosmological model:
\begin{equation}
g(z)=\left(\frac{dz}{dx} \right)x^{(1-\gamma)}F(x).
\label{epseqn}
\end{equation}
Here $x$ is the comoving distance at a redshift $z$ and $F(x)$ is the
curvature factor in the Robertson-Walker metric,
\begin{equation}
ds^2=c^2 dt^2 - a^2[dx^2/F(x)^2+x^2 d\theta^2+x^2 sin^2 \theta
d\phi^2].
\end{equation}   
If the redshift distribution of sources, $dN/dz$, is well-known, then
the amplitude of \wt \ can be predicted for a given power-law model of
\xix, such that measurements of \wt \ can be used to place
constraints on the evolution of \xix.

\subsection{Redshift Distribution as a Function of Magnitude}

In the literature there is a wide variety of adopted redshift
distributions for magnitude-limited samples, which is 
alarming as the predicted amplitude of \wt \ is quite sensitive to the
assumed shape, and particularly the width, of the redshift
distribution.  Most authors use either galaxy evolution models
or small observational samples (N$\sim100$) to construct $dN/dz$ 
distributions.
\cite{McCracken01} use their own galaxy evolution models which are 
tested against an HDF sample of 120 galaxies and which reproduce the
trend of median magnitude with $z$ and dispersion in redshift for a
given magnitude range as seen by the Canada-France Redshift Survey (CFRS).  
\cite{Cabanac00} use the
CFRS luminosity function combined with the PEGASE spectral atlas
\citep{Fioc97} to construct models of the $dN/dz$ distributions.  As
they point out, the choice of which atlas to use can affect the 
K-corrections by as much as 50\%, so that their $dN/dz$ distribution
is quite sensitive to this choice.  \cite{Roche99} use the 
\cite{Charlot96} spectral evolution models combined with the
luminosity function of the Las Campanas Redshift Survey data.

Other authors have used observed redshift distributions from
relatively small, deep spectroscopic surveys.  \cite{Postman98} use
the empirical function $dN/dz \propto z^2 {\rm exp}[-(z/z_0)^2]$, fit to the
CFRS spectroscopic data.  These fits
are presumably based on the sample of $\sim600$ galaxies for which
CFRS obtained secure redshifts, with the highest $z\sim1.3$.  They
also use various evolving Schechter luminosity functions to model
$dN/dz$ and find parameters which provide good fits to their \wt \
results.  \cite{Wilson03} use the form $dN/dz \propto z^2 {\rm
exp}(-z/z_0)$, where the mean redshift is $3z_0$ and the median
redshift is $2.67 z_0$.  They fit for $z_0$ using the SSA22 field
sample of \cite{Cowie99}, which has $\sim200$ galaxies to $I=23.25$, where
corrections are made for faint galaxies without redshifts which are
thought to lie between $z=1.5-2.0$.  This parameterization of $dN/dz$
is markedly different from that of \cite{Postman98}, with more galaxies at
higher redshift, and leads to very discrepant predictions of the amplitude 
of \wt \ as a function of magnitude (at $I_{\rm AB}=23$ the CFRS $dN/dz$ model
predicts a value of $A_\omega$ which is 6\% lower for $\epsilon=0$,
14\% lower for $\epsilon=1$, and 23\% lower for $\epsilon=2$; see Section
5.5 for a definition and discussion of $\epsilon$).

Here we use a uniform sample of 2954 galaxies with secure redshifts in early
DEEP2 Galaxy Redshift Survey \citep{Davis02, Coil03xisp, Madgwick03deep} data
in the Extended Groth Strip (EGS), where redshifts are observed from
$z\simeq0-1.5$.  We find that the form used by \cite{Wilson03},
$dN/dz \propto z^2 {\rm exp}(-z/z_0)$, fits the DEEP2 data reasonably well.
The overall normalization of $dN/dz$ is irrelevant in estimating the
amplitude of \wt; it is the shape, and particularly the width, of
$dN/dz$ that is important for our purposes.  The $dN/dz$
parameterization \cite{Postman98} used for the CFRS data does not fit
the DEEP2 sample; their narrower function contains too few objects at
high redshifts.

\begin{figure}
\centerline{\scalebox{.6}{\includegraphics{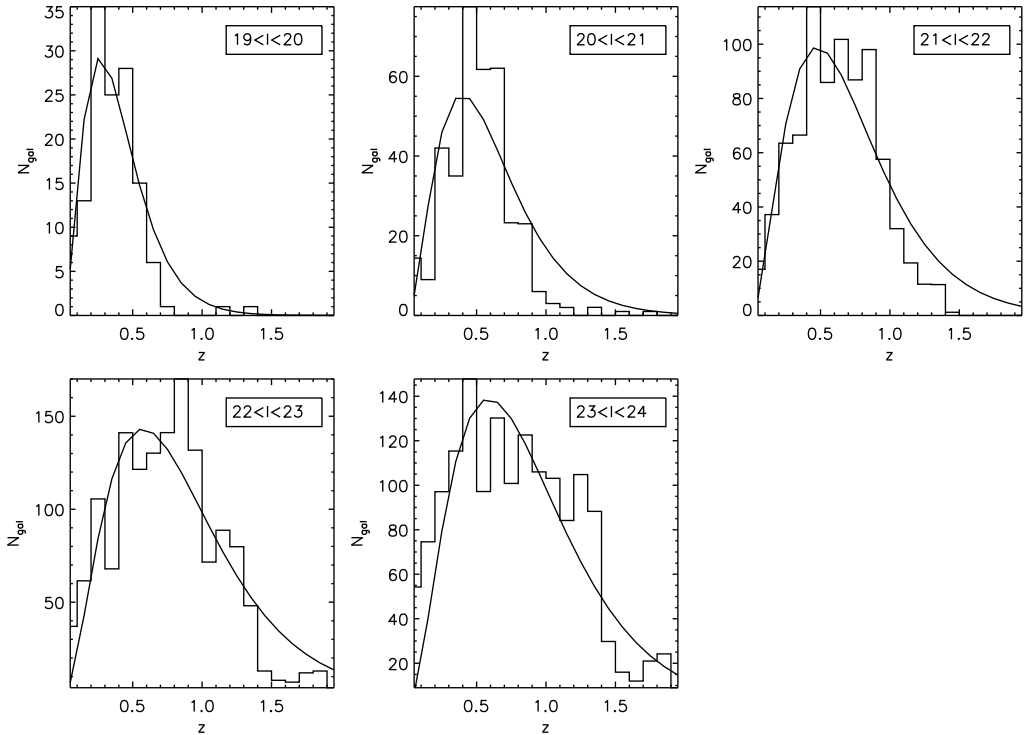}}}
\caption{Redshift histograms of galaxies in the early DEEP2 Redshift Survey
data for independent $I_{\rm AB}$ magnitude ranges. 
Fits for $dN/dz \propto z^2 {\rm exp}(-z/z_0)$ are plotted as 
solid curves, and values for $z_0$ are listed in Table \ref{zdisttable}. 
The mean 
redshift in each magnitude range 
is $3z_0$ and the median redshift is $2.67 z_0$.
\label{zdist_hist}}
\end{figure}

To reconstruct the redshift distribution of sources in the DEEP2 data
we have inversely weighted each galaxy by its probability of
selection, which is a function of magnitude \citep{Newman04}.  Figure
\ref{zdist_hist} shows the redshift distribution of sources in the
DEEP2 data along with our best-fit $dN/dz$ distribution for differential 
magnitude bins.  As discussed in \cite{Coil03xisp}, the DEEP2 Galaxy Redshift
Survey does not measure redshifts for galaxies at $z\geq1.45$ as the
$\lambda$ 3727 \AA \ [OII] doublet and Ca H+K absorption features move
out of the observable spectral wavelength range.  Therefore, any
objects at $z\gtrsim1.45$ are not included in the sample used to derive
$dN/dz$.  This will affect the fainter magnitude ranges in particular,
which will be missing objects at $z\geq1.45$.  Chuck Steidel has
followed up our blue ($(B-R)\leq0.4$) failures using LRIS-B (private
communication) and has found that of the 20 objects he has targeted,
all were at $z>1.45$, with a median $z=1.75$ and a standard deviation
of $\sigma\sim0.30$.  This redshift distribution is very similar to 
that shown in
\cite{Steidel04} for the 'BM' galaxies, which lie at $<z>=1.70 \pm0.34$,
and it seems quite likely that our blue failures are in the `BM' population.  
We correct our redshift
distributions by assigning redshifts to the blue galaxies without
measured redshifts such that they have a Gaussian distribution with a
mean of $z=1.75$ and a standard deviation of $\sigma=0.30$.  Our
resulting redshift distributions are shown in Figure \ref{zdist_hist}.
Fits for $z_0$ for each magnitude range are shown in Table
\ref{zdisttable} with and without the correction for missed galaxies
at $z>1.45$, which affected only the fainter magnitude bins.  Errors
on $z_0$ were calculated using bootstrap resampling of the data and
may underestimate errors due to cosmic variance, as the 2954 redshifts
which we use are from a single $\sim0.5$ x 0.3 degree field.

\begin{figure}
\centerline{\scalebox{.6}{\includegraphics{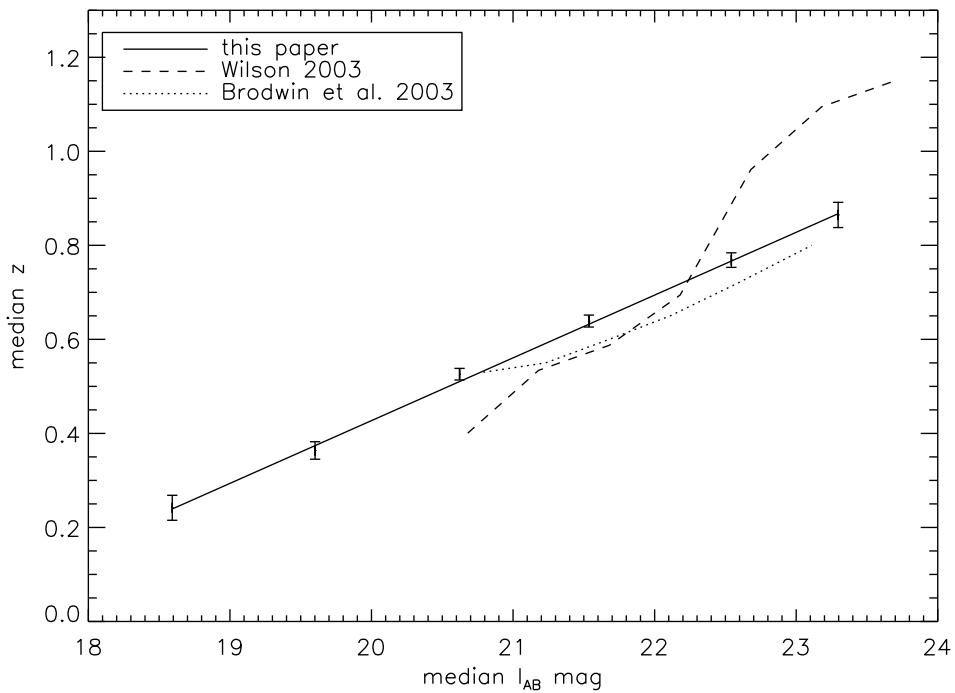}}}
\caption{Median redshift as a function of $I_{\rm AB}$ magnitude for data
from the DEEP2 Redshift Survey, using the fits shown in Figure 
\ref{zdist_hist}, where the median redshift is $2.67 z_0$. The solid line
is a linear fit to the data:  $z_0=-0.84+0.050$ median $I_{\rm AB}$.
Results from two recent surveys are also shown.  \label{medianz}}
\end{figure}

After fitting our $dN/dz$ model to the DEEP2 spectroscopic sample, we
find that $z_0$ is well-fit as a linear function of the median $I_{\rm
AB}$ magnitude: $z_0=-0.84+0.050$ median $I_{\rm AB}$, where the median
redshift is $2.67z_0$ and the mean redshift is $3z_0$.  Figure \ref{medianz}
plots our data points for the median redshift as a function of 
median $I_{\rm AB}$ magnitude, with a solid line showing the linear fit.
We also include recent results from \cite{Wilson03} and \cite{Brodwin03}.
\cite{Wilson03} use the same functional form for $dN/dz$ that we use here,
though their fits of $z_0$ vs magnitude do not match ours well at
magnitudes fainter than $I_{\rm AB}\sim22$, where they find a higher
median redshift than we do.  However, their results are based on a small
sample of $\sim200$ galaxies.  Our results agree better with those 
of \cite{Brodwin03}, though they find systematically 
lower values for the median redshift as a function of magnitude and 
use a narrower $dN/dz$ fitting form which has fewer galaxies at 
higher redshifts.

We also include in Table \ref{z0table} fits for $dN/dz$ in integral
magnitude ranges for both $R_{\rm AB}$ and $I_{\rm AB}$ bands, which
are of interest for other analyses such as weak gravitational lensing
studies.  For the integral magnitude bins we fit for two
distributions, $dN/dz \propto z^2 {\rm exp}(-z/z_0)$ and $dN/dz
\propto z^2 {\rm exp}(-z/z_0)^{1.2}$, and find that they both fit the
data well to within the noise.  For the $z^2 {\rm exp}(-z/z_0)^{1.2}$
distribution, the mean redshift is $2.09z_0$ and the median $1.91z_0$.
Beyond our limiting magnitudes of $R_{\rm AB}=24$ and $I_{\rm
AB}=23.5$, we extrapolate to fainter magnitudes using a simple linear
fit to the $z_0$ values in the brighter bins.  Note that all 
magnitudes are AB in the native CFHT 12k filter system.

\subsection{Constraining Evolution of Clustering}

Given the redshift distribution of sources, one can use measurements
of \wt \ to either determine $\xi(x)$ at a given redshift, if the
galaxies lie in a narrow redshift range, or parameterize the
redshift-dependence of $\xi(x,z)$ and solve for the local value of
$\xi(x,0)$ and its evolution with redshift.

\cite{Groth77} suggested the `$\epsilon$-model' for the evolution of
the two-point correlation function:
\begin{equation}
x_0(z)=x_0(0) (1+z)^{-(3+\epsilon-\gamma)/\gamma}
\end{equation}
\begin{equation}
\xi(x,z)=\left[\frac{x_0(0)}{x} \right]^\gamma
(1+z)^{-(3+\epsilon-\gamma)}
\end{equation}
where $x$ is a comoving distance, $x_0(0)$ is the local
value of $x_0(z)$ 
and $\gamma$ is the slope of $\xi(x)$.  In proper coordinates, 
the scale-length evolves as $r_0(z)=r_0(0) (1+z)^{-(3+\epsilon)/\gamma}$ and 
$\xi(r)$ evolves as $\xi(r,z)=\xi(r,0) (1+z)^{-(3+\epsilon)}$.  This model
was motivated such that $\epsilon=0$ for clustering fixed in proper
coordinates, where $n \xi=$constant and $n$ is the mean galaxy density.  
Clustering which is fixed in comoving coordinates
requires $\epsilon=\gamma-3$, which equals $-1.2$ for $\gamma=1.8$
(equivalent to $\delta=-0.8$).  Any value of $\epsilon>0$ leads to
growth of clustering in proper coordinates.  Linear theory predicts
$\epsilon=\gamma-1$, equal to $0.8$ for $\gamma=1.8$, for growth of
dark matter clustering in an $\Omega_M=1$ cosmology.  Using this
$\epsilon$ parameterization for the growth of clustering, equation
\ref{Aeqn} becomes
\begin{equation}
A=\frac{\int_{0}^{\infty} x_0^\gamma(0) (1+z)^{-(3+\epsilon-\gamma)}
g(z)\left(\frac{dN}{dz}\right)^2
dz}{\left[\int_{0}^{\infty}\left(\frac{dN}{dz} \right) dz \right]^2}.
\end{equation}

\begin{figure}
\centerline{\scalebox{.6}{\includegraphics{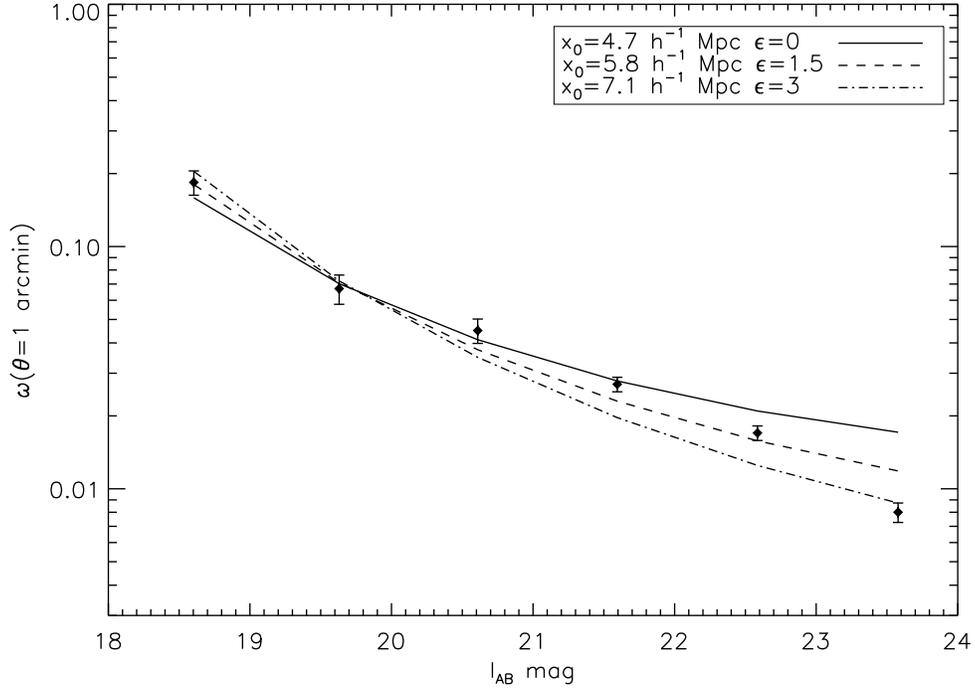}}}
\caption{The amplitude of \wt \ at $1\arcmin$ 
as a function of the median $I_{\rm AB}$ magnitude compared with
predictions for different values of
the local clustering scale-length, $x_0(0)$, and an evolutionary parameter,
$\epsilon$, such that $x_0(z)=x_0(0) (1+z)^{-(3+\epsilon-\gamma)/\gamma}$.
\label{ourmodel}}
\end{figure}

Figure \ref{ourmodel} shows models of the predicted amplitude of \wt \
as a function of magnitude for different values of $x_0(0)$ and
$\epsilon$, given our fits for $dN/dz$.  An increase in $x_0(0)$ has
the effect of shifting the predicted values higher, increasing the
amplitude of clustering at all magnitudes, while increasing $\epsilon$
creates a steeper slope in the evolution of clustering.  We find that
our data is not well-fit by any single value of $\epsilon$; the
brightest galaxies are best-fit by $\epsilon\sim0$ and $x_0(0)\sim5$
\mpch, to $I_{\rm AB}\sim21$.  The amplitude of \wt \ at $1\arcmin$ 
then begins to decrease more rapidly than an $\epsilon=0$ model predicts, 
such that the fainter galaxies imply a value of $\epsilon\sim3$.  This
implies significant growth of clustering at 
intermediate redshifts ($z\sim0.5-1$) 
and clustering fixed in physical coordinates in the more recent past 
($z\sim0-0.5$).
A model with $\epsilon\sim3$ does not match our data within the
errors across the full magnitude range; in particular, the model predictions
are low for $I_{\rm AB}\sim20-23$.  The $\epsilon$ model is known
not to fit angular clustering data well \citep{McCracken01, Wilson03}
and is in fact not a physically realistic model given the presence of
a galaxy bias which evolves with redshift.  Therefore we attempt to
find a new model which better fits the data.

\begin{figure}
\centerline{\scalebox{.6}{\includegraphics{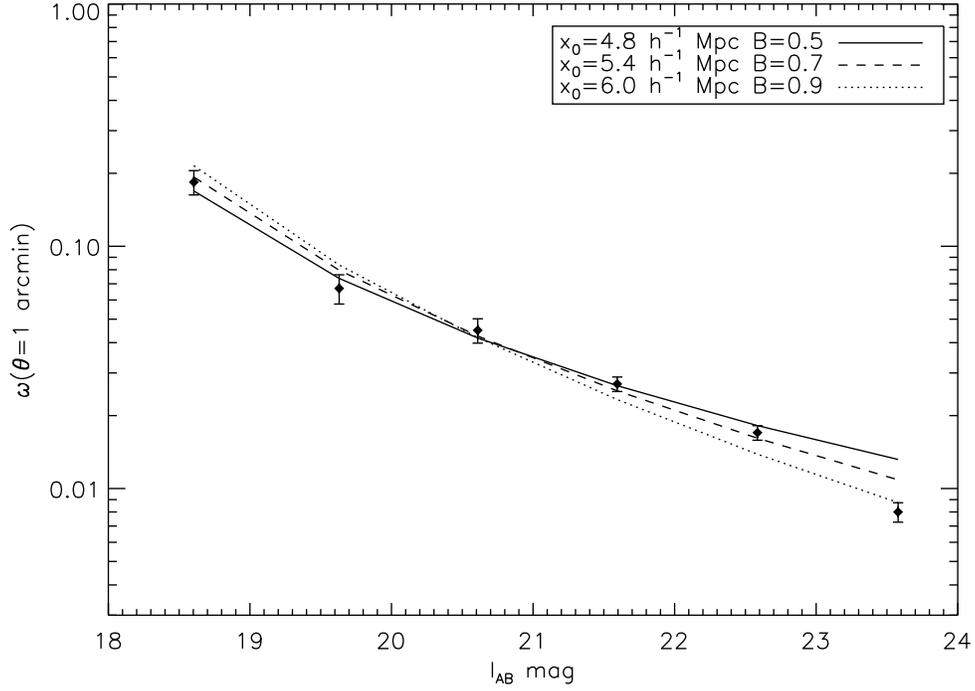}}}
\caption{Same as Figure \ref{ourmodel}, for 
a model in which $x_0(z)$ evolves linearly with redshift: 
$x_0(z)=x_0(0)(1-Bz)$, for $z\sim0-1.5$.  \label{newmodel}}
\end{figure}

Given that \wt \ declines monotonically with the median $I_{\rm AB}$
magnitude of a sample, and that $z_0$ is well-fit as a linear function
of the median $I_{\rm AB}$ magnitude, we propose a linear model for
the evolution of $x_0(z)$, $x_0(z)=x_0(0)(1-Bz)$.  
We stress that this linear model for the evolution of $x_0(z)$ 
is valid only in the redshift range this dataset is sensitive to, 
namely $z\sim0-1.5$, and $x_0$ is expected to increase at higher 
redshifts, as seen in both data and simulations \citep[e.g.,][]{Adelberger03, 
Kauffmann99b}.  For this model, \xir \ evolves as
\begin{equation}
\xi(x,z)=\left[\frac{x_0(z)}{x}\right]^{\gamma}=
\left[\frac{x_0(0)(1-Bz)}{x}\right]^{\gamma}= \xi(x,0)(1-Bz)^{\gamma}.
\end{equation}
Replacing the $x_0(z)$ term in Equation \ref{Aeqn} with
$x_0(0)(1-Bz)$, we can fit for $B$ and $x_0(0)$.
Fits for the linear model are shown in Figure \ref{newmodel}; the best
fit is $x_0(0)\sim5-6$ \mpch \ and $B\sim0.6-0.9$.  These values imply that
$x_0(z=0.8)\sim1.4-3.1$ \mpch \ and $x_0(z=1.1)\sim0.1-2.0$ \mpch. 
\xir \ results from early DEEP2 data measure $x_0(z=0.8)=3.5 \pm0.8$ 
\mpch \ and $x_0(z=1.1)=3.1 \pm0.7$ \mpch \ \citep{Coil03xisp}, 
somewhat higher than what
is implied here, though within 1-2$-\sigma$ of the upper range of values. 

We do not necessarily expect the results to match exactly, as the \xir \
results are for a data sample with a limiting magnitude of $R_{\rm AB}=24.1$,
which is roughly half a magnitude brighter than our faintest \wt \ sample.
Additionally, 
a luminosity-dependent bias, which is expected to exist at $z\simeq1$
 \citep[e.g.,][]{Coil03xisp, Kauffmann99b}, would cause the \wt \ results from
our faintest magnitude bin to imply a somewhat lower value of $x_0(z)$.
There are other possible factors that may contribute to this 
discrepancy; a wider $dN/dz$ distribution, especially at the higher redshifts,
would imply larger values of $x_0(z)$, though comparisons to other surveys do
not indicate that our redshift distributions are too narrow.  
Limber's equation also has implicit assumptions which may 
break down at some level; 
for example, Limber's equation holds that the probability of finding
a galaxy pair at a given separation depends on \xir \ multiplied by
the selection function for each galaxy.  This implies that galaxies have a 
selection function which is independent of their clustering.  This assumption 
does not hold if clustering depends on environment or luminosity, for example,
which is known to be true both locally and at intermediate redshifts.  This
may affect our ability to interpret \wt \ results when using
a sample that includes a mix of galaxy types, luminosities and redshifts.

\section{Galaxy Clustering as a Function of $(R-I)$ Color}

\begin{figure}
\centerline{\scalebox{.6}{\rotatebox{90}{\includegraphics{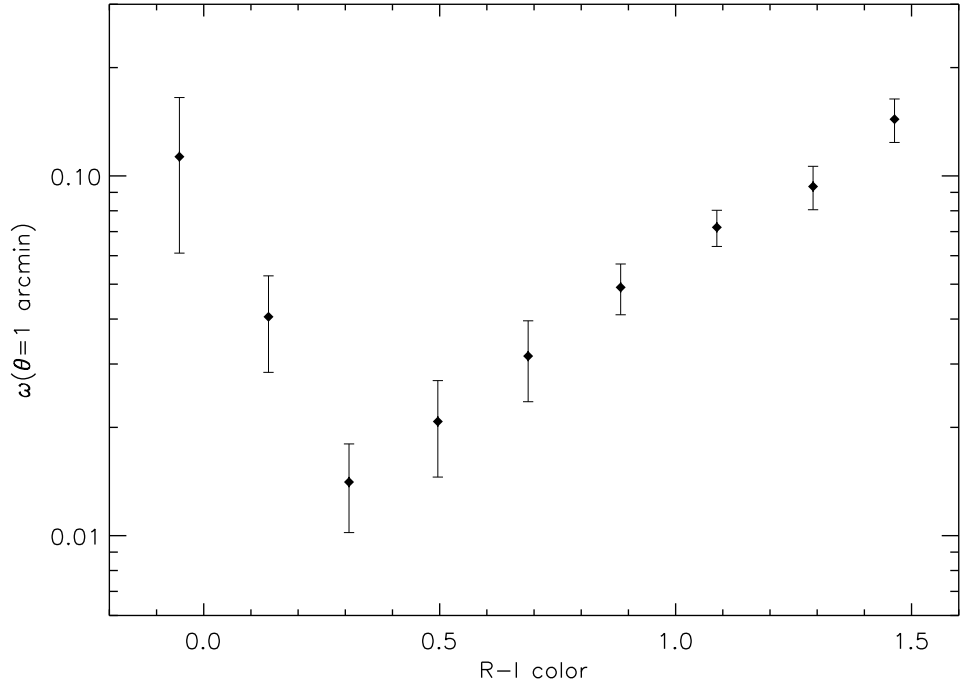}}}}
\caption{The amplitude of \wt \ measured at 1$\arcmin$ shown as a 
function of $(R-I)$ color, where the slope has been measured simultaneously (see Figure \ref{colorslope}).  \label{color}}
\end{figure}

We now parallel the previous section, discussing the 
clustering as a function of color instead of magnitude.  
We plot in Figure \ref{color} the amplitude of \wt \ at $1\arcmin$ 
for nine independent $(R-I)$ color bins, where we fit for the slope and 
amplitude
simultaneously.  Each bin has a width of $(R-I)=0.2$, and we use all
galaxies in the magnitude range $I_{\rm AB}=18-24$.  Table
\ref{colortable} lists the computed amplitude and errors of \wt \
measured at 1$\arcmin$ for each color range.  Similar to \wt \ as a function
of magnitude (Section 5.2), 
we find that the results are consistent if we first sum the 
DD, DR, and RR 
pairs across all the pointings before taking the ratio in Equation (2) and
errors are estimated using bootstrap resampling of the data.
There is a strong,
monotonic progression between $0.2<(R-I)<1.5$ of redder galaxies having a
higher clustering strength.  This trend is not surprising, as red
galaxies have long been known to be more strongly clustered than blue
galaxies locally and this trend appears to continue to 
$z\simeq1$ (recent results include Willmer et al. 1998, Zehavi et al. 
2002, Coil et al. 2003)\nocite{Willmer98, Zehavi02, Coil03xisp}.  
We also find that there is
a sharp rise in the clustering amplitude for the bluest galaxies in our
sample ($-0.2<(R-I)<0.0$), which have a clustering strength comparable to
the reddest galaxies ($1.2<(R-I)<1.6$).  
\cite{Landy96} first reported such a marked correlation of clustering
with color, in a sample of $\sim3000$ galaxies selected by $U-R_F$
color, where they found that the bluest and reddest galaxies showed
much strong clustering than the sample as a whole.  More recently,
with larger surveys, \cite{McCracken01} and
\cite{Wilson03} also find similar clustering trends with $V-I$ color 
and report that the bluest and reddest samples are highly clustered. 
We note that the strong rise in clustering amplitude seen for the bluest
galaxies is only apparent after rather stringent color cuts are made 
($R-I<0.2$), and that galaxies which are selected to be blue based on
a $B$ magnitude alone do not show strong clustering 
\citep[e.g.,][]{Efstathiou91}.

\begin{figure}
\centerline{\scalebox{.6}{\rotatebox{90}{\includegraphics{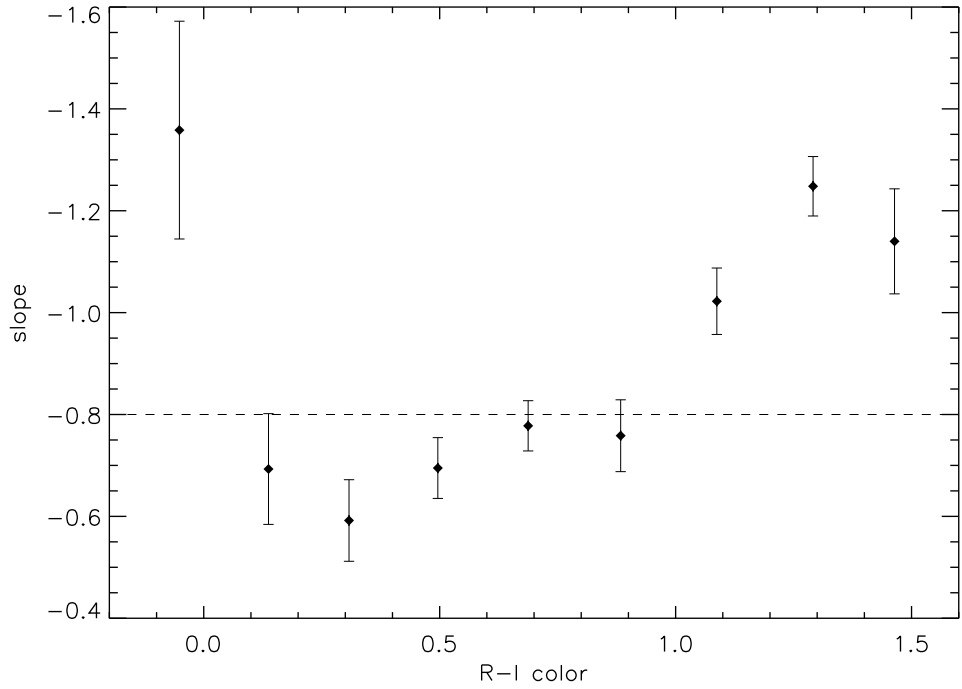}}}}
\caption{The best-fit slope, $\delta$, of \wt \ as a function of $(R-I)$ color.
 \label{colorslope}}
\end{figure}

Figure \ref{colorslope} shows the best-fit slope for each color bin,
as listed in Table \ref{colortable}.  We find that galaxies with
$0.2<(R-I)<1.0$ have slopes of $-0.6\lesssim\delta\lesssim-0.8$, while the 
bluest and
reddest samples have steeper slopes, with $-1.0\lesssim\delta\lesssim-1.4$.  
This
implies that the slope of $\xi(r)$ for the bluer and redder samples is
steeper than for galaxies with $0.2<(R-I)<1.0$.  Locally, the slope of \xir \
in SDSS galaxies is found to be $\gamma=1.86 \pm0.03$ for redder 
galaxies with a restframe color of $u^*-r^*>1.8$ and $\gamma=1.41 \pm0.04$ 
for bluer galaxies with $u^*-r^*<1.8$ \citep{Zehavi02}.  At intermediate
redshifts ($0.1< z < 0.5$) \cite{Shepard01} find that galaxies
with early-type SEDs have a clustering slope of $\gamma=1.91 \pm0.06$ while
late-type galaxies show $\gamma=1.59 \pm0.08$.  Our results imply a slope
of $\gamma\gtrsim2.0$ for the reddest galaxies, and $\gamma\sim1.6-1.7$ for
typical blue galaxies.  The very bluest galaxies, 
with $(R-I)<0$, appear not to 
be typical intermediate-redshift blue galaxies, as described in 
the following sections.

\subsection{Redshift Distributions of Different Color Ranges}

Using the sample of 2954 early DEEP2 galaxies with secure redshifts 
between $0<z<1.5$ in
the EGS we construct redshift distributions for sources as a function
of $(R-I)$ color as shown by the solid lines in Figure \ref{colorzhist}.
Note that these histograms do not have the same scale on the y-axis,
as the number of galaxies in each bin varies widely.  The reddest
galaxies, with $1.2<(R-I)<1.6$, lie in a relatively narrow redshift
range centered near $z\sim0.85$.  Successively bluer bins have both a
lower median redshift and a wider redshift distribution.  Galaxies
with $0.0<(R-I)<0.4$ appear to be predominately centered at $z\sim0.35$, though
there is a tail to higher redshifts.  The bluest galaxies
($-0.2<(R-I)<0.0$) have a narrow redshift distribution between
$z\sim0-0.5$.  As discussed above, the DEEP2 Redshift Survey can not
measure redshifts for galaxies at $z\gtrsim1.45$ as observable emission
and absorption line features move out of the spectral wavelength range
of our data.  These redshift histograms are therefore missing galaxies
at $z\gtrsim1.45$.  The redshift success rate of the DEEP2 survey in
the EGS is fairly insensitive to color in the range $0.2<(R-I)<1.6$,
where the success rate is $\geq80\%$ and where 90\% of the targeted
objects lie, but it begins to drop blueward of $(R-I)\sim0.2$.  For
galaxies with $0.0<(R-I)<0.2$ the redshift success rate is $\sim55\%$, while
the bluest galaxies, with $-0.2<(R-I)<0.0$, have a success rate of
$\sim30\%$. We believe that most of the observed galaxies without
redshifts lie at $z>1.45$, as these blue galaxies should have emission
lines which we would detect at lower redshifts.  The sample of 20
DEEP2 blue galaxies without measured redshifts which Chuck Steidel has
followed were all found to lie at $z>1.45$ (see Section 5.4 for details).  
We therefore correct the
bluest two redshift distributions by assigning redshifts to the
galaxies without measured redshifts such that they have a Gaussian
distribution with a mean of $z=1.75$ and a standard deviation of
$\sigma=0.30$, matching the distribution of sources followed by
Steidel.  These corrected redshift distributions can been seen in
Figure \ref{colorzhist} as the dashed lines for the two bluest
distributions.

\subsection{Clustering Scale-length as a Function of $(R-I)$ Color}

\begin{figure}
\centerline{\scalebox{.6}{\includegraphics{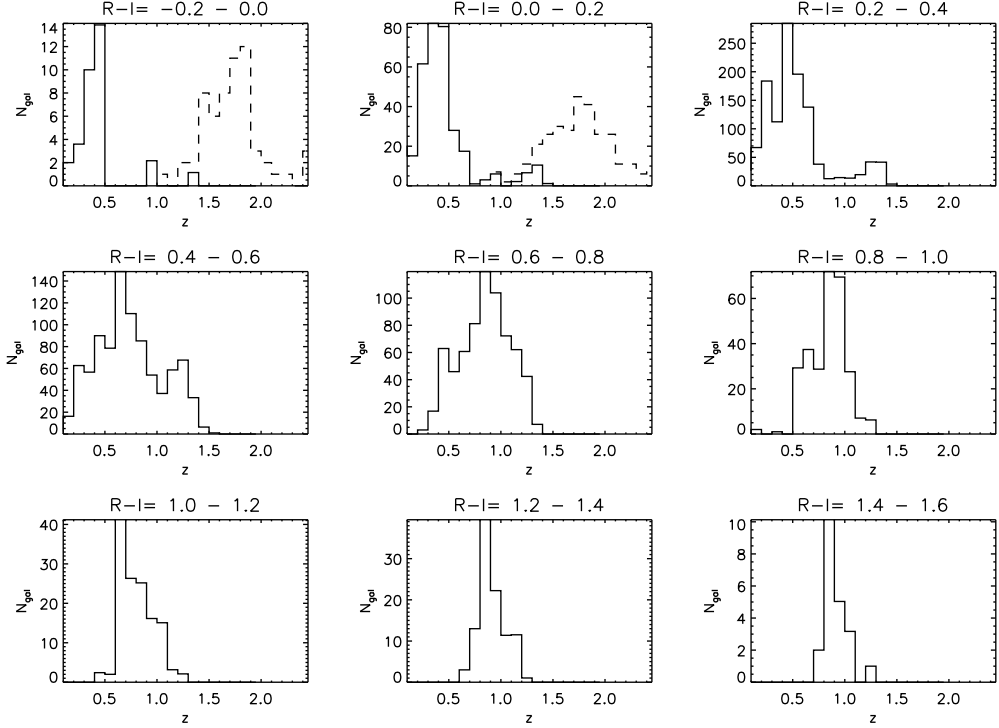}}}
\caption{Redshift distribution of DEEP2 EGS sources as a function of $(R-I)$ color.  The dashed lines for the two upper left plots show the assumed 
redshift 
distribution for galaxies at $z>1.45$ (see text for details).
 \label{colorzhist}}
\end{figure}

\begin{figure}
\centerline{\scalebox{.6}{\rotatebox{90}{\includegraphics{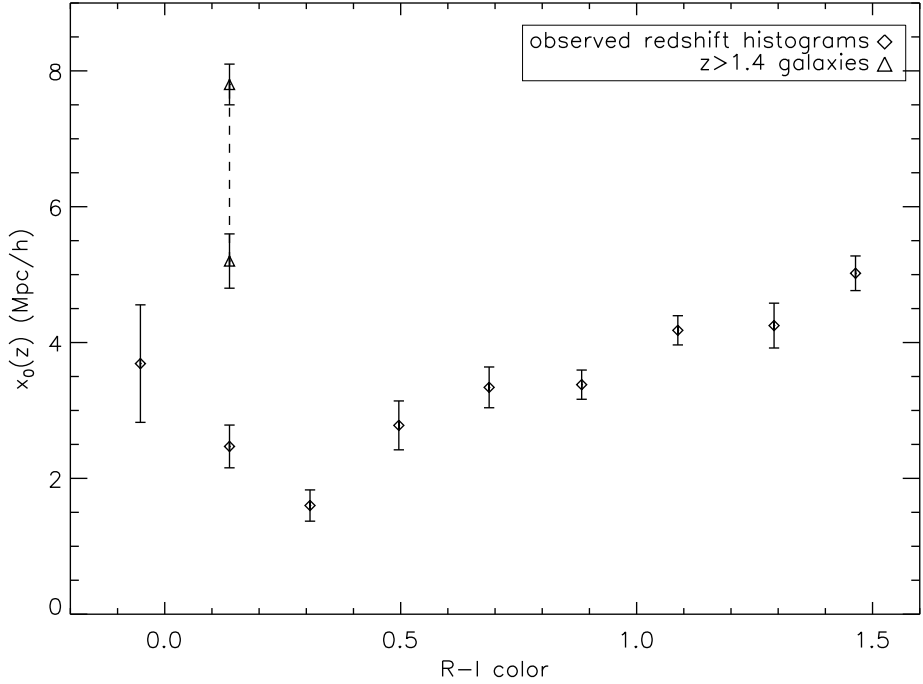}}}}
\caption{The comoving scale-length $x_0(z)$ as a function of observed
  $(R-I)$ color.  In addition to the scale-length derived using the
  observed redshift distributions shown in Figure \ref{colorzhist}, we
  also plot values of the scale-length derived for $z\sim1.7$ galaxies
  (see text for details). 
 \label{colorr0}}
\end{figure}

Figure \ref{colorr0} shows the comoving scale-length $x_0$ as a
function of $(R-I)$ color calculated using the observed redshift
distributions shown in Figure \ref{colorzhist} in solid lines,
assuming that $x_0$ is constant over the redshift range of galaxies in
each color bin.  The errors on $x_0$ are derived from the errors on
\wt \ alone and do not include uncertainties in the redshift
distribution; therefore they may be underestimates of the true error.
Values of $x_0$ and associated errors are given in Table
\ref{colortable}.  As dividing the galaxies into color bins results in
more uniform populations within a single bin, we believe that Limber's
equation should be more robust in interpreting \wt \ as a function of
color.

For the reddest galaxies, with $1.4<(R-I)<1.6$, we find $x_0=5.02
\pm0.26$ \mpch \ (note that this is the comoving scale-length at the redshift
of the sample).  These galaxies, centered at $z\sim0.85$, have
absorption-dominated spectra and are likely to be massive 
progenitors of local ellipticals.  Our value is somewhat 
lower than that found by \cite{Brown03}, who
measure early-type galaxies with $B_W-R>1.28$ between $0.7<z<0.9$ to
have $x_0=6.6 \pm0.8$ \mpch.  Locally, \cite{Budavari03} finds from an
angular clustering study of SDSS galaxies that $x_0(0)=6.59 \pm0.17$
\mpch \ and $\delta=-0.96 \pm0.05$ for early-type galaxies with
$-21>M_r>-23$.  Our results at $z\sim0.85$ are consistent with 
predictions of little evolution of $x_0(z)$ for early-type galaxies
from $z\sim1$ to $0$ \citep{Kauffmann99b, Benson01}.  
Our reddest sample 
may be a less-extreme version of the extremely red object (ERO)
population which are seen to have large clustering at $z\sim1$
with $x_0 \sim7.5-12 $ \mpch \ \citep[e.g.,][]{Daddi01,Firth02}.

The comoving scale-length decreases as galaxies become bluer to a
minimum of $x_0=1.60 \pm0.23$ \mpch \ at $0.2<(R-I)<0.4$.  These galaxies lie
predominately between $z\sim0.3-0.6$, though there is a small tail at
$z>1.0$.  The redshift distribution of galaxies with $0.0<(R-I)<0.2$ is
strongly bimodal; this sample appears to be a mix of two populations: relatively
local galaxies
similar to the sample with $0.2<(R-I)<0.4$ and a $z>1.4$ population
of bright, star-forming galaxies.  If the angular clustering is due
entirely to the lower-redshift population, then $x_0=2.47 \pm0.32$ \mpch,
where we have used the uncorrected redshift distribution shown as a
solid line in Figure \ref{colorzhist}.  If the clustering is due
entirely to the higher-redshift population, then $x_0=3.8$ \mpch.  More
likely, however, is that the observed \wt \ value is a mixture of these 
two populations, and we would like to separate their contributions 
to \wt \ and measure values of $x_0$ for each population, as they
may have different intrinsic clustering properties.  It can be shown
that \wt \ from a combined sample of disparate populations with different
clustering properties and non-overlapping $dN/dz$ distributions 
adds linearly such that
\begin{equation}
\omega(\theta)_{total}=f_1^2 \omega_1(\theta) + (1-f_1)^2 \omega_2(\theta)
\end{equation}
where $f_1$ is the fraction of the total number of galaxies in one
population, $(1-f_1)$ is the fraction in the other population, and
$\omega_1(\theta)$ and $\omega_2(\theta)$ are the angular clustering
amplitudes of each population \citep[e.g.,][]{Meiksin99}.  
The value of $x_0=3.8$ \mpch \ which
we find for the $z>1.4$ population is therefore a lower limit on 
the clustering scale-length of that population, as
some of the observed projected clustering is 
due to the lower-redshift sample.  If we use
the value of \wt \ observed for the $0.2<(R-I)<0.4$ sample as the
clustering contribution from the lower-redshift population, we can
then subtract this from the observed \wt \ amplitude for the
$0.0<(R-I)<0.2$ sample, assuming that $\sim$50\%-75\% of the galaxies 
in this color bin are in the lower-redshift population.  
Using the redshift distribution of the
$z>1.4$ galaxies shown in Figure \ref{colorzhist}, 
we find $x_0=7.8 \pm0.3$ \mpch \ for the $z>1.4$ galaxies if 50\% of
the galaxies in this color bin are in this population 
(corresponding to all the galaxies without measured redshifts), and 
$x_0=5.2 \pm0.4$ \mpch \ if 25\% of the galaxies 
are at $z>1.4$ (corresponding to half of the galaxies without
measured redshifts).  These values most likely bracket the true value, as 
the percentage of galaxies without measured redshifts at $z>1.4$ is
likely to be between 50\%-100\%.  These results are shown in Figure \ref{colorr0}.  
It is worth noting that this error does not
include uncertainties in the redshift distribution of $z>1.4$
galaxies.

In the bluest color bin, $-0.2<(R-I)<0.0$, the number of galaxies
drops dramatically and the redshift distribution for objects at
$z<1.4$ appears to be relatively narrow, though the robustness of 
this width is uncertain given the small sample size.  If the clustering
of the bluest objects is due to galaxies with the uncorrected redshift
distribution shown as a solid line in the upper left panel of Figure
\ref{colorzhist}, then $x_0=3.69 \pm0.87$ \mpch.  However, as the
redshift success rate is only $\sim30\%$ for galaxies in this bin in
the DEEP2 EGS region, it is likely that most of the objects in this
color range are at $z>1.4$, since they must be actively forming stars
to have such blue colors and would therefore have strong nebular
emission lines which would allow DEEP2 to determine their redshifts
easily if at $z<1.4$.  A third population also contributes to objects found in 
this color bin: broad-line AGN found in the DEEP2 sample disproportionally have
$(R-I)<0$.  In the EGS, where the photometric-redshift pre-selection
of targets is not applied, 11\% of the galaxies in the bluest bin are
identified as broad-line AGN, whereas $\lesssim1$\% are objects with
$(R-I)>0$ are so classified.  These percentages are lower limits,
since the spectroscopic signatures of weak AGN could be dominated by
galaxy light and therefore remain undetected.  The AGN identified in
DEEP2 spectra span the full redshift range $0<z<1.5$.  The bluest
color bin, $-0.2<(R-I)<0.0$, thus includes at least three distinct
populations: faint blue galaxies at low redshift, bright star-forming
galaxies at $z>1.4$, and broad-line AGN with strong, blue continua.

Separating the contributions to the observed total \wt \ 
from each population is not entirely straightforward, as there are 
uncertainties in both the estimates of the fraction of objects
in each population and their redshift distributions.
However, given the small percentage of local galaxies seen in the
observed redshifts for this bin and the low clustering amplitude of the
local star-forming galaxies in the $0.2<(R-I)<0.4$ bin, which have 
$x_0=1.60 \pm0.23$ \mpch, it seems unlikely that local galaxies 
alone account for the large clustering signal seen here.
A sample of local galaxies with the same relatively redshift 
distribution and clustering 
scale-length as those in the $0.2<(R-I)<0.4$ color range, would have
a \wt \ value 13\% as large as what is measured in this bin, 
in opposition to the conclusions of \cite{Landy96}, \cite{McCracken01}
and \cite{Wilson03}, who suggest that the strong clustering of blue
galaxies is dominated by a local galaxy population.

If the observed \wt \ in this color bin was due
entirely to $z>1.4$ galaxies, the clustering scale-length for that
population would be $x_0=7.5 \pm1.5$ \mpch.  If AGN were the
sole contributing population and had the same redshift distribution
as the local galaxies in this color bin, they would have a clustering
scale-length of $x_0=3.69 \pm0.87$ \mpch, the same as for local galaxies.
Observations of AGN and QSO clustering have found 
scale-lengths comparable to local early-type galaxies, with $x_0\sim6-8$
\mpch, and very little evolution in the scale-length with
redshift \citep{Croom02, Grazian04}.  It therefore seems quite feasible 
that AGN contribute significantly to the large clustering seen in this color
range.  Assuming that the observed \wt \ is due to a combination of these three 
populations, and that each of them is totally independent, would lead 
to much larger values for $x_0$ of each 
subsample, possibly in excess of other estimates of the clustering of 
$z>1.4$ galaxies and AGN.  It is likely that not only the
autocorrelation of each population, but also its cross-correlation with 
the other sets of extremely blue objects, leads to the strong observed 
clustering in this color bin.

In order to check the degree of independence between the redshift 
distributions in each sample, we measure the cross-correlation 
between galaxies in color bins $i$ and $j$, $w_{ij}(\theta)$, 
and normalize each correlation amplitude, $A_{w_{ij}}$, 
by the square root of the auto-correlation function amplitude, $A_{w_{ii}}$:
\begin{equation}
c_{ij}=\frac{A_{w_{ij}}(1\arcmin)}{\sqrt{A_{w_{ii}}(1\arcmin)A_{w_{jj}}(1\arcmin)}}
\end{equation}
where we have measured $w(1\arcmin)$ assuming a constant slope of 
$\delta=-0.8$ and $w_{ij}$ is defined by equation \ref{dd} 
where DD is replaced
with D$_i$D$_j$.  This results in a covariance matrix, shown
in Table \ref{covar}, that 
 provides information on the physical covariance between 
galaxies in each color bin.  Our results are generally consistent with
the covariance of the redshift distributions seen in Figure \ref{colorzhist}.
Each color sample shows significant covariance with the immediately adjacent
samples, and galaxies in the reddest four color bins, with $0.8<(R-I)<1.6$, 
all have 
significant covariance with each other.  The bluest color bin does not 
have significant covariance with galaxies with $0.2<(R-I)<0.4$, implying that
the relatively local, blue, star-forming galaxies with low correlation
seen in the $0.2<(R-I)<0.4$ range contribute very little to the clustering
seen in the bluest population, as we assumed above.
However, the $z>1.4$ galaxy population must contribute to both of the
bluest two color bins.  This covariance matrix provides a reassuring check
on our redshift distributions used to interpret \wt \ as a function of color.

\section{Summary and Conclusions}

We measure the projected angular correlation function, \wt, using deep
photometric imaging with a sample of
350,000 galaxies to $I_{\rm AB}=24$, a significant subset of which is
used as photometric data for the DEEP2 Galaxy Redshift Survey. 
 Our data covers a total of
 5 degrees$^2$ over 5 separate fields, a larger area than
has been studied by previous deep angular clustering surveys.  
This increase in area, combined
with galaxy redshift distributions derived from a relatively large, uniform
spectroscopic sample of 2954 DEEP2 galaxies between $0<z<2$, leads to 
 tighter constraints on models 
than has been previously available.  Using robust redshift
distributions as a function of color also allows for quantitative  
interpretations of the strong trends seen in \wt \ vs. observed $(R-I)$.

We find that there has been
significant growth of clustering from $z>1$ to $z\sim0$, and that the
`$\epsilon$-model' proposed by \cite{Groth77} does not describe the
evolution of clustering well.  We propose an alternative model in which the
comoving scale-length, $x_0(z)$, evolves linearly over the redshift
range we are sensitive to here and find that this model better 
describes the data.  We
find that $x_0\sim1-3$ \mpch \ at $z\sim1$, though these results are
not straightforward to interpret, as they result not only from
intrinsic evolution, but also any changes in the mean luminosity, color, and
type of galaxies with redshift.

We also find that \wt \ is a very sensitive function of
observed $(R-I)$ color, with both the reddest and bluest galaxies
showing strong clustering amplitudes and steeper clustering slopes.
Red galaxies with $1.4<(R-I)<1.6$ have a comoving scale-length of
$x_0=5.02 \pm0.26$ \mpch \ and lie in a narrow redshift range centered at
$z\sim0.85$.  Their spectra indicate that these galaxies have older
stellar populations and are likely to be progenitors of local
ellipticals.  The clustering amplitude decreases with color to a minimum at
$0.2<(R-I)<0.4$, where $x_0=1.60 \pm0.23$ \mpch \ for blue star-forming 
galaxies between $z\sim0.3-0.6$.  The bluest galaxies, with $(R-I)<0$, are
very highly clustered, and include additional populations of 
$z>1.4$ bright, star-forming galaxies with a scale-length of $x_0\gtrsim5$ 
\mpch \ and broad-line AGN.  We find that the clustering of
the bluest galaxies is not likely to be dominated by a local
population or due entirely to high-redshift galaxies.

\
We thank Chuck Steidel, Jim Peebles and David Hogg for useful discussions.
We also thank the DEEP2 team for generously providing early redshift data 
and Andy Connolly for his vital work on obtaining photometric calibrations 
for the DEEP2 fields.  An anonymous referee provided insightful and
constructive comments.
This project was supported by the NSF grant AST-0071048.  
J.A.N. acknowledges support by NASA through Hubble Fellowship grant
HST-HF-01165.01-A awarded by the Space Telescope Science Institute, which is
operated by AURA Inc. under NASA contract NAS 5-26555.  
C.-P. M is partially supported by a Cottrell Scholars Award from the
Research Corporation and NASA grant NAG5-12173.

\

\begin{deluxetable}{ccccccc}
\tabletypesize{\small}
\tablewidth{0pt}
\tablecaption{Details of our CFHT12k observations.}
\tablehead{
\colhead{Field} & 
\colhead{pointing} & 
\colhead{RA} & 
\colhead{Dec.} & 
\colhead{exp. time (hrs)} &
\colhead{seeing} & 
\colhead{date of obs.} \\ 
\colhead{} & \colhead{} & \colhead{} & \colhead{} & 
\colhead{B \ \ R \ \ I} & 
\colhead{(FWHM)} & \colhead{}}
\startdata
0230 & 1 & 02 27 30 & 00 35 00 & 1.0 \ 1.0 \ 2.2 & 0.8 $\arcsec$ &
     10-11/99 09/00 \\ & 2 & 02 30 00 & 00 35 00 & 0.7 \ 1.0 \ 2.0 &
     0.7 $\arcsec$ & 10-11/99 09/00 \\ & 3 & 02 32 30 & 00 35 00 &
     0.8 \ 1.0 \ 2.0 & 0.7 $\arcsec$ & 10-11/99 09/00 \\
1052 & 1 & 10 52 43 & 56 58 48 & 1.0 \ 1.5 \ 1.0 & 0.69 $\arcsec$ &
04/01 \\
1416 & 1 & 14 15 29 & 52 08 19 & 1.0 \ 1.0 \ 1.0 & 0.9 $\arcsec$ &
     05/00 04/01 \\ & 2 & 14 17 55 & 52 35 07 & 1.8 \ 1.0 \ 1.3 &
     1.0 $\arcsec$ & 05/00 04-05/01 \\ & 3 & 14 20 21 & 53 01 55 &
     1.2 \ 1.5 \ 2.0 & 0.7 $\arcsec$ & 05/00 04-05-07/01 \\
1651 & 1 & 16 48 00 & 34 55 02 & 1.7 \ 2.5 \ 1.0 & 0.9 $\arcsec$ &
     05-09-10/00 05/01 \\ & 2 & 16 51 30 & 34 55 02 & 1.8 \ 1.0 \
     1.0 & 0.9 $\arcsec$ & 05-09/00 04-05/01 \\ & 4 & 16 55 28 & 33
     42 00 & 1.0 \ 1.8 \ 1.7 & 0.8 $\arcsec$ & 05-09-10/00 \\
2330 & 1 & 23 26 56 & 00 08 00 & 1.3 \ 1.0 \ 3.3 & 0.8 $\arcsec$ &
     10-11/99 09/00 \\ & 2 & 23 30 00 & 00 08 00 & 1.0 \ 1.0 \ 3.0 &
     0.8 $\arcsec$ & 10-11/99 09/00 \\ & 3 & 23 33 03 & 00 08 00 &
     1.0 \ 1.0 \ 2.0 & 0.7 $\arcsec$ & 10-11/99 09/00 \\ & 4 & 23 25
     40 & 00 -40 00 &0.8 \ 1.0 \ 0.5 & 0.8 $\arcsec$ & 10/99 09/00 
     \\ & 5 & 23 34 00 & 00 50 00 & 1.2 \ 1.0 \ 1.0 & 0.8 $\arcsec$ &
     10/99 09-10/00 \\
\enddata
\label{obstable}
\end{deluxetable}

\begin{deluxetable}{cccccc}
\tablewidth{0pt}
\tablecaption{Power law fits to \wt$=A_{\omega}\theta^{\delta}$ as a function of $I_{\rm AB}$ magnitude.}
\tablehead{
\colhead{$I_{\rm AB}$ range} & \colhead{median $I_{\rm AB}$} & \colhead{\# gals/pt.} & 
\colhead{\onew} & \colhead{\onew} & \colhead{$\delta$} \\
&&& \colhead{($\delta=-0.8$)}  && }
\startdata
$18-19$ & 18.6 & 152  & $0.187 \pm0.023$ & $0.184 \pm0.027$ & $-0.82 \pm0.10$ \\
$19-20$ & 19.6 & 453  & $0.070 \pm0.010$ & $0.067 \pm0.010$ & $-1.01 \pm0.09$ \\
$20-21$ & 20.6 & 1224 & $0.060 \pm0.004$ & $0.045 \pm0.005$ & $-0.99 \pm0.05$ \\
$21-22$ & 21.6 & 2812 & $0.031 \pm0.001$ & $0.027 \pm0.002$ & $-0.90 \pm0.04$ \\
$22-23$ & 22.6 & 5921 & $0.015 \pm0.001$ & $0.017 \pm0.001$ & $-0.74 \pm0.05$ \\
$23-24$ & 23.6 & 11879 & $0.007 \pm0.001$ & $0.008 \pm0.001$ & $-0.70 \pm0.07$ \\
\enddata
\label{magtable}
\end{deluxetable}

\begin{deluxetable}{cccc}
\tablewidth{0pt}
\tablecaption{Fits for the redshift distribution of DEEP2 
spectroscopic sources as a function of magnitude, for differential
magnitude bins,
 assuming $dN/dz \propto z^2 {\rm exp}(-z/z_0)$.  
The corrected values of $z_0$ include galaxies at $z>1.4$ 
(see text for details).}
\tablehead{
\colhead{$I_{\rm AB}$ range} & \colhead{uncorrected $z_0$} & \colhead{corrected $z_0$} 
& \colhead{$z_0$ error}}
\startdata 
18 - 19 & 0.091 & 0.091 & 0.010 \\
19 - 20 & 0.136 & 0.136 & 0.007 \\
20 - 21 & 0.197 & 0.197 & 0.005 \\
21 - 22 & 0.237 & 0.239 & 0.005 \\
22 - 23 & 0.278 & 0.288 & 0.006 \\
23 - 24 & 0.285 & 0.324 & 0.010 \\
\enddata
\label{zdisttable}
\end{deluxetable}

\clearpage

\begin{deluxetable}{lccclcc}
\tablecaption{Fits for the redshift distribution of DEEP2 
spectroscopic sources as a function of magnitude, for integral
magnitude bins.}
\tablehead{ \colhead{$R_{\rm AB}$ range} & \colhead{$z_0$}\tablenotemark{a} & \colhead{$z_0$}\tablenotemark{b} & \colhead{} & \colhead{$I_{\rm AB}$ range} & \colhead{$z_0$}\tablenotemark{a} & \colhead{$z_0$}\tablenotemark{b}}
\startdata
18 - 20 & 0.104 & 0.145 & \ & 18 - 20 & 0.123  &  0.171 \\
18 - 21 & 0.136 & 0.188 & \ & 18 - 21 & 0.168  &  0.232 \\
18 - 22 & 0.168 & 0.232 & \ & 18 - 22 & 0.206  &  0.284 \\
18 - 23 & 0.207 & 0.285 & \ & 18 - 23 & 0.242  &  0.332 \\
18 - 24 & 0.255 & 0.350 & \ & 18 - 24\tablenotemark{c} & 0.283  &  0.389 \\
18 - 25\tablenotemark{c} & 0.286 & 0.392 & \ & 18 - 25\tablenotemark{c} & 0.322  &  0.443 \\
18 - 26\tablenotemark{c} & 0.324 & 0.443 & \ & 18 - 26\tablenotemark{c} & 0.362  &  0.496 \\
18 - 27\tablenotemark{c} & 0.361 & 0.494 & \ & 18 - 27\tablenotemark{c} & 0.401  &  0.550 \\
\enddata
\label{z0table}
\tablenotetext{a}{for $dN/dz \propto z^2 {\rm exp}(-z/z_0)$}
\tablenotetext{b}{for $dN/dz \propto z^2 {\rm exp}(-z/z_0)^{1.2}$}
\tablenotetext{c}{extrapolated using a linear fit (see text for details)}
\end{deluxetable}

\begin{deluxetable}{cccccccc}
\tabletypesize{\small}
\tablewidth{0pt}
\tablecaption{Power law fits to \wt$=A_{\omega}\theta^{\delta}$ 
as a function of $(R-I)$ color, using the observed redshift histograms shown in Figure 16 (solid lines).}
\tablehead{
\colhead{$(R-I)$} & \colhead{median} & \colhead{\# gals/pt} 
& \colhead{\onew} & \colhead{\onew} & \colhead{$\delta$} & 
\colhead{$x_0(z)$} \\
\colhead{range} & \colhead{$(R-I)$} & & \colhead{($\delta=-0.8$)} & & & \colhead{Mpc $h^{-1}$} \\}
\startdata
$-0.2-0.0$ & $-0.05$ & 274 & $0.118 \pm0.021$ & $0.113 \pm0.023$ & $-1.36 \pm0.21$ & $3.69 \pm0.87$ \\ 
 $0.0-0.2$ & 0.14   & 1869 & $0.037 \pm0.003$ & $0.041 \pm0.005$ & $-0.69 \pm0.11$ & $2.47 \pm0.32$ \\ 
 $0.2-0.4$ & 0.31   & 5796 & $0.012 \pm0.001$ & $0.014 \pm0.001$ & $-0.59 \pm0.08$ & $1.60 \pm0.23$ \\
 $0.4-0.6$ & 0.50   & 5745 & $0.019 \pm0.001$ & $0.021 \pm0.001$ & $-0.69 \pm0.06$ & $2.78 \pm0.36$ \\
 $0.6-0.8$ & 0.69   & 4415 & $0.031 \pm0.001$ & $0.032 \pm0.002$ & $-0.78 \pm0.05$ & $3.34 \pm0.30$ \\
 $0.8-1.0$ & 0.88   & 2364 & $0.048 \pm0.003$ & $0.049 \pm0.003$ & $-0.76 \pm0.07$ & $3.38 \pm0.22$ \\
 $1.0-1.2$ & 1.09   & 1221 & $0.080 \pm0.005$ & $0.072 \pm0.006$ & $-1.02 \pm0.07$ & $4.18 \pm0.22$ \\
 $1.2-1.4$ & 1.29    & 861 & $0.098 \pm0.006$ & $0.093 \pm0.006$ & $-1.25 \pm0.06$ & $4.25 \pm0.33$ \\
 $1.4-1.6$ & 1.46    & 308 & $0.171 \pm0.014$ & $0.144 \pm0.016$ & $-1.14 \pm0.10$ & $5.02 \pm0.26$ \\
\enddata
\label{colortable}
\end{deluxetable}


\begin{deluxetable}{cccccccccc}
\tablewidth{0pt}
\tabletypesize{\small}  
\tablecaption{Covariance of \wt \ as a function of $(R-I)$ color}
\tablehead{
\colhead{$(R-I)$} & \colhead{-0.2-0.0} & \colhead{0.0-0.2} & \colhead{0.2-0.4} & \colhead{0.4-0.6} & \colhead{0.6-0.8} & \colhead{0.8-1.0} & \colhead{1.0-1.2} & \colhead{1.2-1.4} & \colhead{1.4-1.6} \\
\colhead{range} & \colhead{} & \colhead{} & \colhead{} & \colhead{} & \colhead{} & \colhead{} & \colhead{} & \colhead{} & \colhead{} \\}
\startdata
-0.2-0.0 & 1.00 &   0.66 &   0.14 &  -0.06 &  -0.22 &  -0.19 &  -0.06 &  -0.14 &  -0.36 \\
 0.0-0.2 & 0.66 &   1.00 &   0.62 &   0.12 &  -0.13 &  -0.16 &  -0.19 &  -0.20 &  -0.26 \\
 0.2-0.4 & 0.14 &   0.62 &   1.00 &   0.64 &   0.26 &   0.09 &  -0.06 &  -0.01 &  -0.02 \\
 0.4-0.6 &-0.06 &   0.12 &   0.64 &   1.00 &   0.70 &   0.38 &   0.22 &   0.25 &   0.10 \\
 0.6-0.8 &-0.22 &  -0.13 &   0.26 &   0.70 &   1.00 &   0.72 &   0.57 &   0.63 &   0.43 \\
 0.8-1.0 &-0.19 &  -0.16 &   0.09 &   0.38 &   0.72 &   1.00 &   0.87 &   0.71 &   0.49 \\
 1.0-1.2 &-0.06 &  -0.19 &  -0.06 &   0.22 &   0.57 &   0.87 &   1.00 &   0.92 &   0.54 \\
 1.2-1.4 &-0.14 &  -0.20 &  -0.01 &   0.25 &   0.63 &   0.71 &   0.92 &   1.00 &   0.81 \\
 1.4-1.6 &-0.36 &  -0.26 &  -0.02 &   0.10 &   0.43 &   0.49 &   0.54 &   0.81 &   1.00 \\
\enddata
\label{covar}
\end{deluxetable}

\clearpage

\end{document}